\documentclass[a4paper,fleqn,usenatbib]{mnras}

\usepackage{newtxtext,newtxmath}

\usepackage[T1]{fontenc}
\usepackage{ae,aecompl}

\usepackage{graphicx}	
\usepackage{amsmath}	
\usepackage{amssymb}	

\newcommand{\Msun}{{\mathrm M}_{\odot}}
\newcommand{\nH}{n_{\mathrm{H}}}
\newcommand{\ccm}{\,\mathrm{cm}^{-3}}
\newcommand{\Zsol}{{\mathrm Z}_{\odot} }
\newcommand{\kms}{\,\mathrm{km\,s}^{-1}}

\title[TDGs in cosmological simulations]{Tidal dwarf galaxies in cosmological simulations}

\author[S. Ploeckinger et al.]{
Sylvia~Ploeckinger,$^{1}$\thanks{E-mail: ploeckinger@strw.leidenuniv.nl}
Kuldeep~Sharma,$^{1}$
Joop~Schaye,$^{1}$
Robert A. Crain,$^{2}$ 
\newauthor Matthieu Schaller,$^{3}$ and
Christopher Barber$^{1}$
\\
$^{1}$Leiden Observatory, Leiden University, PO Box 9513, NL-2300 RA Leiden, the Netherlands\\
$^{2}$Astrophysics Research Institute, Liverpool John Moores University, 146 Brownlow Hill, Liverpool L3 5RF, UK\\
$^{3}$Institute for Computational Cosmology, Durham University, South Road, Durham, DH1 3LE, UK
}

\date{Accepted XXX. Received YYY; in original form ZZZ}

\pubyear{2017}

\begin{document}
\label{firstpage}
\pagerange{\pageref{firstpage}--\pageref{lastpage}}
\maketitle

\begin{abstract}
The formation and evolution of gravitationally bound, star forming substructures in tidal tails of interacting galaxies, called tidal dwarf galaxies (TDG), has been studied, until now, only in idealised simulations of individual pairs of interacting galaxies for pre-determined orbits, mass ratios, and gas fractions. Here, we present the first identification of TDG candidates  in fully cosmological simulations, specifically the high-resolution simulations of the EAGLE suite. The finite resolution of the simulation limits their ability to predict the exact formation rate and survival timescale of TDGs, but we show that gravitationally bound baryonic structures in tidal arms already form in current state-of-the-art cosmological simulations. In this case, the orbital parameter, disc orientations as well as stellar and gas masses and the specific angular momentum of the TDG forming galaxies are a direct consequence of cosmic structure formation. We identify TDG candidates in a wide range of environments, such as multiple galaxy mergers, clumpy high-redshift (up to $z=2$) galaxies, high-speed encounters, and tidal interactions with gas-poor galaxies.
We present selection methods, the properties of the identified TDG candidates and a roadmap for more quantitative analyses using future high-resolution simulations.
\end{abstract}




\section{Introduction}

When gas-rich disk galaxies interact, the resulting tidal forces can lead to the formation of extended arms of gas, dust, and stars. Gaseous over-densities in tidal arms that become self-gravitating can collapse, form stars \cite[e.g.][]{lisenfeld_molecular_2016}, and result in kinematically bound structures with baryonic masses in the same range as dwarf galaxies \citep{barnes_formation_1992}. 

The formation of such ``tidal dwarf galaxies" (TDGs) has been observed in many interacting galaxies starting with \citet{mirabel_genesis_1992} and \citet{duc_young_1998}.  For more observations see also e.g. \citet{weilbacher_tidal_2000} for a sample of TDGs in 10 interacting galaxies, \citet{mendes_de_oliveira_candidate_2001} for 7 TDGs in Stephan's Quintet, \citet{kaviraj_tidal_2012} for a comprehensive study of the TDGs in the Sloan Digital Sky Survey, and \citet{lee-waddell_frequency_2016} for 3 TDGs in the two interacting groups NGC 4725/47 and NGC 3166/9. 

TDGs are most easily identified when they are located within a tidal arm and their stellar population is still young and therefore brightest. Their later evolutionary stages are more difficult to observe, although \citet{duc_identification_2014} detected TDG candidates with spectroscopic ages of up to 4 Gyr around the early type galaxy NGC5557. This was the first observational indication that some TDGs can survive their early evolutionary phase when they are actively star-forming and turn into long-lived fossil TDGs. Recently, \citet{kemp_tidal_2016} investigated the tidal filament of NGC 4660 and found faint peaks in the surface brightness. They speculate that two of these peaks could be evolved TDGs that are even older than the ones detected in the tail of NGC5557, but deeper observations are necessary to confirm this. Analytic \citep{ploeckinger_tides_2015} and numerical studies \citep[e.g.][]{bournaud_tidal_2006, recchi_early_2007, yang_reproducing_2014, ploeckinger_chemo-dynamical_2014, ploeckinger_chemodynamical_2015} support the potential longevity of TDGs.

As TDGs form out of disk material from their host galaxy and their masses are too low to capture the high velocity dispersion dark matter particles of the more massive haloes of their host galaxies, they cannot contain a significant amount of dark matter \citep{barnes_formation_1992, bournaud_tidal_2006, wetzstein_dwarf_2007}.
On the contrary, hierarchically formed dwarf galaxies are dominated by dark matter in the current cosmological standard model ($\Lambda$CDM, lambda/dark energy and cold dark matter) as their baryonic matter first condenses in dark matter haloes and later grows by merging with other dark matter-dominated structures and smooth accretion. 

The expected number density of fossil TDGs in a cosmological volume depends on the TDG formation rate and their average survival timescale, both of which are potentially redshift-dependent. The estimates of the fraction of dwarf galaxies that could have a TDG origin span a wide range: from 6\% \citep{kaviraj_tidal_2012}, 10\% \citep{bournaud_tidal_2006}, 16\% \citep{sweet_choirs_2014}, 50\% \citep{hunsberger_formation_1996}\footnote{Note that some of the TDGs identified by \citet{hunsberger_formation_1996} are undetected in narrow-band H$\alpha$ observations and are most likely background sources \citep{eigenthaler_star_2015}.}, to even 100\% \citep{okazaki_dwarf_2000} depending on the underlying assumptions.

As both the galaxy merger rate and the gas fraction of the participating galaxies increases with redshift, the formation rate of objects formed in gas-rich encounters is expected to be redshift dependent. The properties of TDGs that form at high redshift and their further evolution are however largely unknown. Do they get disrupted and contribute to the stellar halo of their host galaxy? At low redshift interacting galaxies either form compact star clusters \citep[e.g.][]{mullan_star_2011} or extended TDGs in their tidal arms. Are both channels present for high redshift galaxy encounters? If young, massive star clusters build up today's globular cluster population, what do the high-redshift TDGs evolve into? Because of their low surface brightness, observations of the formation of TDGs beyond the local volume are not yet possible, therefore numerical simulations are necessary to address these questions. 

Early simulations from \citet{toomre_galactic_1972} already illustrated that the extent and mass of the tidal arms depends on the geometry of the interaction as well as the rotation orientation of the galactic disks (prograde or retrograde encounter). Recently, \citet{barnes_transformations_2016} studied the relation between the formation of tidal tails and the encounter geometry in more detail. Similar to studies of tidal features in galaxy interactions, the formation of TDGs has been investigated in idealised scenarios, typically containing two initially isolated (unperturbed) galaxy disks on parabolic orbits with set mass ratios and gas mass fractions. For example, \citet{bournaud_tidal_2006} produced a set of 96 N-body simulations of interacting galaxies, varying their relative encounter velocity, impact parameter, orbit inclination, orbit orientation (prograde/retrograde), and mass ratios and identified favourable conditions for TDG formation. 

Galaxy mergers in cosmological simulations of representative volumes are fully self-consistent in the sense that the properties of the galaxies (e.g. mass, gas fraction, morphological type, metallicity) as well as the geometry of their encounters (e.g. peri-centre distance, ellipticity, disk orientation, prograde/retrograde rotation) are a natural outcome of their cosmic history and their surrounding large-scale structure, and are not free parameters. More complicated systems with more than two interacting galaxies are included accordingly. Studying TDGs directly in cosmological boxes rather than in idealised galaxy interactions is therefore the next logical step to constrain their formation rates and compare their properties to observations, but until recently this was not possible due to insufficient resolution. Here we use the high-resolution runs of the EAGLE suite of cosmological simulations \citep{schaye_eagle_2015, crain_eagle_2015} with initial gas particle masses of $10^5\Msun$ which enables us to resolve the mass of typical TDGs (around $10^8\Msun$) with several hundreds of particles. This enables for the first time the identification of TDGs in a cosmological sample of interacting galaxies over a large redshift range. 

As the minimum gravitational force softening in these simulations is 0.35 proper kpc, we cannot distinguish between compact star clusters with a half mass radius of a few pc and extended TDGs, both of which are observed in tidal arms \citep[for a review see e.g.][]{duc_tides_2013}. 
We therefore refer to all identified structures that fulfil the TDG criteria as TDG candidates or TDGC(s).

The paper is structured as follows: In Sec.~\ref{sec:method} we describe the cosmological simulations and estimate the expected number of TDGCs by calculating the galaxy merger rates as well as the selection criteria for the (Sec.~\ref{sec:stepselection}) TDG candidates. The results are summarised in Sec.~\ref{sec:results} and a thorough discussion of the limitations of the current simulations for the study of TDG and similar anti-hierarchically formed structures can be found in Sec.~\ref{sec:discussion}. We summarize our findings in Sec.~\ref{sec:conclusions}.
Throughout the paper we denote proper lengths or distances with the prefix p (e.g. pkpc, pMpc) and comoving lengths with the prefix c (e.g. ckpc, cMpc).

\begin{table}
\caption{Redshift and lookback times in Gyr for the EAGLE snapshots considered for the TDGC selection.}
\label{tab:snaps}
\begin{center}
\begin{tabular}{lcc}
\hline
Snapshot nr. & redshift & Lookback time [Gyr]\\
\hline
14&2.24&10.89\\
15&2.01&10.56\\
16&1.74&10.07\\
17&1.49&9.51\\
18&1.26&8.88\\
19&1.00&7.97\\
20&0.87&7.36\\
21&0.74&6.70\\
22&0.62&5.99\\
23&0.50&5.23\\
24&0.37&4.13\\
25&0.27&3.25\\
26&0.18&2.32\\
27&0.10&1.35\\
28&0.00&0.00\\
\hline
\end{tabular}
\end{center}
\end{table}

\section{Method}\label{sec:method}
\subsection{The EAGLE simulations}
We study the TDGC population in the high-resolution runs of the EAGLE suite of cosmological simulations \citep{schaye_eagle_2015, crain_eagle_2015}. The EAGLE simulations use a heavily modified version of the N-body Tree-PM smoothed particle hydrodynamics (SPH) code Gadget \citep{springel_cosmological_2005}.
The simulations assumed a $\Lambda$CDM cosmology with parameters based on the 2013 Planck data release: $\Omega_{\mathrm{M}}$ = 0.307, $\Omega_{\Lambda}$= 0.693,$\Omega_{\mathrm{b}}$ = 0.048, $h$ = 0.6777, $\sigma_{\mathrm{b}}$ = 0.8288 and $n_s$ =0.9611 \citep{planck_collaboration_planck_2014}. The package that contains the modifications of the SPH scheme (e.g. pressure-entropy SPH, energy injection) is referred to as Anarchy (Dalla Vecchia et al. in prep.) and its key ingredients are summarized in appendix A of \citet{schaye_eagle_2015}. A more detailed description as well as comparison studies with standard SPH can be found in \citet{schaller_eagle_2015}.

The simulation imposes a polytropic equation of state of the form $P_{\mathrm{EOS}} \propto \rho^{4/3}$ normalised to $T_{\mathrm{EOS}}$ = 8000 K at  $\nH= 0.1\ccm$. This pressure floor ensures that the Jeans scales are marginally resolved for all gas densities \citep{schaye_relation_2008,dalla_vecchia_simulating_2012} and prevents artificial clumping. Stars form stochastically if the gas temperature is within 0.5 dex of the equation of state and the gas density exceeds a metallicity dependent density threshold \citep[for details see][]{schaye_eagle_2015}. Stellar feedback is implemented stochastically as a way to overcome the overcooling problem \citep[e.g.][]{katz_cosmological_1996} that occurs when thermal energy is injected into an unresolved multiphase ISM \citep{dalla_vecchia_simulating_2012}. While EAGLE produces realistic galaxy sizes and stellar masses and therefore does not suffer from severe overcooling, the stochastic feedback might explain why some EAGLE galaxies have holes in their \ion{H}{i} disks that are a factor of several larger than observed \citep{bahe_distribution_2016}. Black hole seeds with masses of $m_{\mathrm{BH}} = 10^5\Msun /h$ are placed in the centre of every halo with a total mass greater than $10^{10}\Msun /h$ that does not already contain a black hole. These seeds grow by gas accretion or mergers and feedback from active galactic nuclei is implemented analogously to stellar feedback \citep[see][for details]{schaye_eagle_2015}.

We examine the formation of TDGs in two simulations, starting from identical initial conditions, of a cubic volume of side 25 cMpc, realised with 752$^3$ particles of collissionless dark matter and an initially equal number of gas particles. The (initial) particle masses are $2.26\times10^5\Msun$ for baryons and $1.21\times10^6\Msun$ for dark matter and the maximum softening length is 350 ppc. The two simulations adopt slightly different values of the parameters governing the efficiency of feedback processes. \mbox{RefL0025N0752} uses parameters calibrated to yield the observed galaxy stellar mass function and the mass-size relation at $z=0.1$ in the largest box (100 cMpc). \mbox{RecalL0025N0752} adopts parameters calibrated to yield a good match to these same diagnostics in simulations with the higher resolution we appeal here \citep[for a discussion on weak and strong convergence see][]{schaye_eagle_2015}. The main difference between the two boxes is that in RefL0025N0752  there are up to twice as many galaxies with stellar masses of around $10^{9}\Msun$ as observed at $z = 0.1$ while RecalL0025N0752 is consistent with the observed number density.

The aim of the work presented here is to provide a proof of concept study of the qualitative formation mechanism of TDGs and a roadmap for future simulations, rather than a prediction of the formation rate of TDGs. A more quantitative analysis is not yet possible due to the limitations in resolution (see Sec.~\ref{sec:discussion} for a more detailed discussion). For this reason both boxes (RefL025N0752 and RecalL025N0752) are considered for this study.

Structures in EAGLE are identified as following: first, dark matter particles are grouped into haloes by a friends-of-friends (FoF) algorithm and baryonic particles (gas and stars) are assigned to the same FoF halo as their nearest dark matter neighbour. The {\textsc{subfind}} algorithm \citep{springel_populating_2001, dolag_substructures_2009} identifies in a subsequent step locally over-dense regions based on isodensity contours around saddle points in density. Finally, {\textsc{subfind}} removes all particles within the over-dense region that are not gravitationally bound, based on their total energy. The result of the algorithm is a therefore a list of gravitationally bound objects (subhaloes) within each FoF halo.
For TDGs and other dark matter free structures, it is crucial that this step includes all particle types, as any algorithm that would define subhaloes based solely on dark matter particles would fail to find potential TDG candidates.

For gas particles, the binding energy of individual particles has to exceed not only the kinetic, but also the thermal energy of the gas \citep{dolag_substructures_2009} to be attributed to a subhalo. This is especially important for TDGs as for them gas particles are the dominant particle type. Without a halo of dark matter particles, the term ``subhalo" might be misleading, but as it is a widely used term for the results of structure finding algorithms in cosmological simulations, we also refer to the identified TDGCs as ``subhalos" throughout the paper. Note, that limiting the sample to self-bound objects is a conservative approach, as we exclude TDGs that are transient features based on their internal kinematics, as suggested for the TDGC in the Antennae galaxies (NGC 4038/39) \citep{hibbard_high-resolution_2001}.

A large number of properties of the identified substructures, the galaxies within them, as well as the FoF haloes are publicly available in the EAGLE database \citep{mcalpine_eagle_2016} for 29 output times (``snapshots") between redshifts 20 and 0 (see Table~\ref{tab:snaps}). \citet{qu_chronicle_2017} link the EAGLE subhaloes across snapshots by tracing their most bound particles and construct merger trees. Each subhalo/galaxy can be identified in one snapshot by the identifier of its FoF halo (``GroupNumber") in combination with its subhalo identifier within the FoF halo (``SubGroupNumber"). A special case in each FoF halo is the subhalo with number 0, which is the central galaxy of this group. All baryonic particles in a FoF halo that are not gravitationally bound to any other subhalo are assigned to subhalo 0 \citep[called the ``background halo" in][]{springel_populating_2001}. Across snapshots the evolution of galaxies can be followed by their unique identifiers (``GalaxyID") and the unique identifiers of their main descendants (``DescendantID") which is the GalaxyID of the subhalo in the next snapshot that contains the largest fraction of the most bound particles of the galaxy that was originally selected \citep[for more details see][]{qu_chronicle_2017}. This means that subhaloes with the same DescendantIDs will merge into one gravitationally bound subhalo. 

\begin{figure} 
	\begin{center}
		\includegraphics*[width=\linewidth]{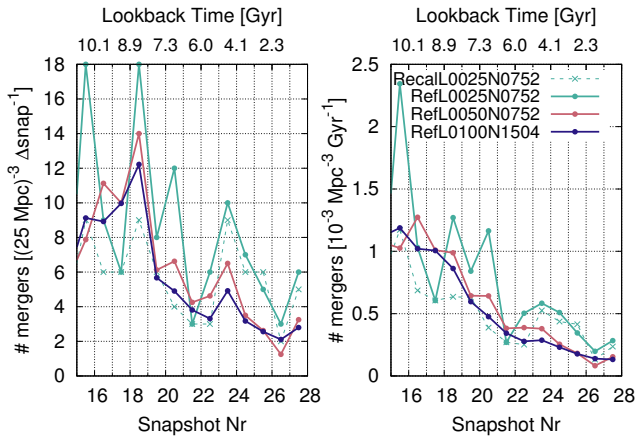}
		\caption{Left: Absolute number of selected galaxy mergers ($M_{\mathrm{baryon}} > 10^9\Msun$, mass ratio $> 1:10$, see text for details) normalized to a 25 cMpc box. Right: As in the left panel but in units of $10^{-3}$ cMpc$^{-3}$ Gyr$^{-1}$. }
		\label{fig:merger_rates}
	\end{center}
\end{figure}

\subsection{Merger rates in EAGLE}\label{sec:estimateTDGnumber}

For a more thorough analysis of mass assembly in EAGLE in general, we refer to \citet{qu_chronicle_2017}. Here, the EAGLE database is used to estimate the number of galaxy merger events between two snapshots that could potentially form a TDG. From snapshot 15 ($z=2$) to 27 ($z=0.1$) galaxy pairs that will merge into the same descendant galaxy were selected. The query was limited to baryonic masses greater than $10^9 \Msun$ for the less massive galaxy and a baryonic mass ratio larger than 1:10. We use the stellar and gas masses within a spherical aperture of 30 pkpc to exclude the mass of extended gaseous and stellar haloes of high-mass galaxies \citep[see][]{schaye_eagle_2015}. For comparison with the larger boxes, we only consider stellar masses of $\log_{\mathrm{10}} M_{*}/\Msun < 10.7$ as the galaxy stellar mass function is incomplete in the 25 Mpc box for higher masses \citep[see][]{schaye_eagle_2015}. The merger rate evolves smoothly from redshift 2 to 0 for the larger box sizes (50 and 100 Mpc) while for the smallest boxes (25 Mpc) the behaviour is more stochastic (Fig.~\ref{fig:merger_rates} right panel) due to the small number of galaxy mergers in the full box (Fig.~\ref{fig:merger_rates} left panel). Both high-resolution simulations (RefL0025N0752 and RecalL0025N0752) trace the merger rates of the larger boxes well, but the total numbers are very low, with the re-calibrated simulation (RecalL0025N0752) including slightly fewer mergers than the reference simulation. Note that the exact merger rate for a given redshift could potentially depend on the halo finding algorithm (see Sec.~\ref{sec:discussion} for a discussion).

The criteria for identifying the hosts of the TDGCs adopted here are not very strict. \citet{bournaud_tidal_2006} found that interactions with mass ratios between 4:1 and 1:8 favour the formation of long-lived TDGs and simulations by \citet{wetzstein_dwarf_2007} highlight the importance of a significant gas content for triggering the collapse of an overdense structure within a tidal arm into a TDG. As {\textsc{subfind}} can significantly underestimate the mass ratios during major mergers (see Sec.~\ref{sec:discussion}), it is not advisable to restrict the mass ratio too much at this step, even though the problem can be reduced by using aperture gas and stellar masses, as is done here.

From Fig.~\ref{fig:merger_rates}  we conclude that, depending on the number of TDGs formed per merger event, the maximum number of TDGs that are expected to form in each snapshot in the simulated cosmological box could be very low (less than a few).  

\subsection{Selecting TDG candidates}\label{sec:stepselection}

In a first step, TDG candidates are selected by querying the EAGLE database for sub-haloes between redshifts 2 and 0 that do not have any dark matter or black hole particles associated with them ($M_{\mathrm{TDGC,DM}}=0$, $M_{\mathrm{TDGC,BH}}=0$) and that have gas masses above $M_{\mathrm{TDG,gas}}$. The fiducial minimum TDG gas mass here is $M_{\mathrm{TDGC,gas}} > 10^7 \,\Msun$ and therefore very close to the resolution limit of the simulation (see Sec.~\ref{sec:discussion} for a discussion). As star formation in EAGLE is implemented stochastically and is not fully sampled in low-mass structures such as TDGs, we only require the stellar mass to be non-zero, which yields a minimum stellar mass of $ M_{\mathrm{TDGC,\star}} > 2.26 \times 10^5\,\Msun$ (one star particle) in the fiducial database query. As we want to use the offset with respect to the stellar mass-metallicity relation as an independent diagnostic quantity in a later step, we do not include TDGCs that consist of gas particles alone. This means that some objects could be missing purely because they stochastically do not form a single star particle while another object with the same properties does. The number of identified TDG candidates is therefore a lower limit, not only because of the possibly unresolved gas collapse (see the discussion on the artificial pressure floor in Sec.~\ref{sec:discussion}), but also because of the poorly sampled stellar population.

In addition to the criteria above, the selected TDG candidates are required to be located between distances of $d_{\mathrm{TDGC-host}} > 2 \times R_{\mathrm{h,gas}}$ and $d_{\mathrm{TDGC-host}} < \mathrm{min}(20 \times  R_{\mathrm{h,gas}}, 200\,\mathrm{pkpc})$ from another galaxy whose gas mass is at least 10 times higher than the TDG candidate or has $M_{\mathrm{host,gas}} > 10^{9}\,\Msun$. Here $R_{\mathrm{h,gas}}$ is the radius that encloses half the gas mass of the subhalo of the more massive galaxy. Especially for central galaxies $R_{\mathrm{h,gas}}$ is an inaccurate measure of the disk size as it includes all gas bound to the subhalo. The additional limit on the distance $d_{\mathrm{TDGC-host}}$ of 200 pkpc ensures that TDGCs are still located in the vicinity of the host galaxy.  This citerion is motivated by the assumption that the resolution is inadequate to follow the evolution of TDGCs over a long time out to larger distances from their host galaxies. 

The minimum distance between a TDGC and its host galaxy is important to avoid the misidentification of self-bound substructures within a galactic disk as TDG candidates. While it is possible to miss the population of TDGs that form close to the disk and fall back on a short timescale, the most long-lived TDGs are expected to form close to the tip of the tidal arm \citep[e.g.][]{bournaud_tidal_2006} and therefore clearly outside the disk of their host galaxy. The chosen parameter values are as liberal as possible because TDGs should also be identified in situations that were not accessible before, such as TDG formation at higher redshifts or in more complicated interaction geometries. In the next step their tidal origin is verified and from the objects that are excluded we can learn about the limitations of selecting TDG candidates purely based on the subhalo properties stored in the EAGLE database.

\begin{table}
\caption{Overview of the TDGC selection parameters.}
\label{tab:parameter}
\begin{center}
\begin{tabular}{lrr}
							& $\ge$							&$\le$  \\
\hline
$M_{\mathrm{TDGC,gas}}$		& $10^7\Msun$ 					&  - \\
$M_{\mathrm{TDGC,\star}}$		&  $2.26\times10^5\Msun$			&  - \\
$M_{\mathrm{TDGC,DM}}$		& 0					&  0 \\
$M_{\mathrm{TDGC,BH}}$		&  0					&  0 \\
$d_{\mathrm{TDGC-host}}$ 		&  $2 \times R_{\mathrm{h,gas}}$		& min($20 \times R_{\mathrm{h,gas}}$, 200 pkpc) \\
$\bar{d}_{\mathrm{TDGC-host,tb}}$ 	&  	-							& min($2 \times R_{\mathrm{h,gas}}$, 70 pkpc)\\
$M_{\mathrm{host,gas}}$			& $10^9\Msun$ 					&  - \\
\hline
\end{tabular}
\end{center}
\end{table}

\begin{figure*} 
	\begin{center}
		\includegraphics*[width=\linewidth]{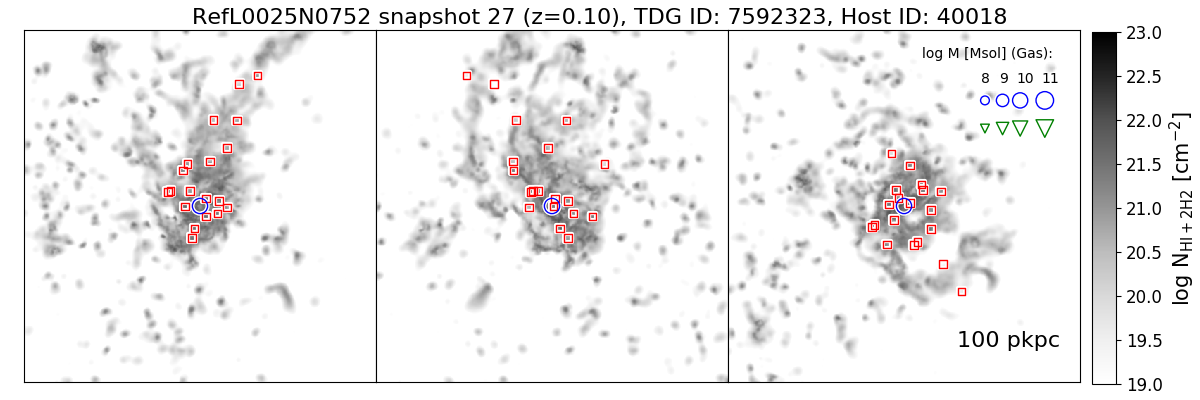}
		\includegraphics*[width=\linewidth]{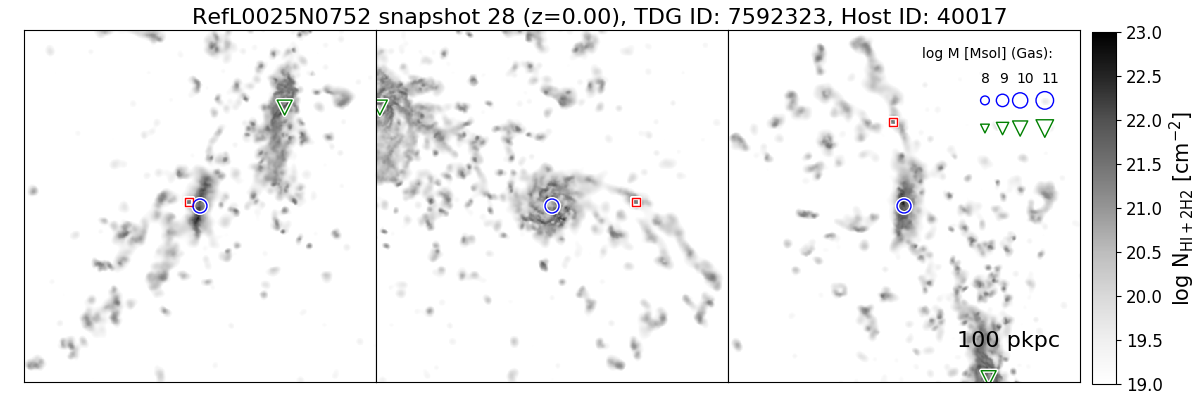}
		\caption{TDGC Ref-1: Neutral hydrogen column density along the line of sight in greyscale for all particles that belong to the same FoF halo as the TDGC and its host galaxy for three different projections along the simulation axes (y-x, z-x, y-z from left to right). The bottom row presents the gas distribution for the snapshot in which the candidate is selected and the top row shows the previous snapshot. The red squares indicate the position of the TDGC in the bottom row  and in the top row they mark the positions of the individual traced back TDGC particles. The blue circle shows the centre of potential of the subhalo that contains the host galaxy. Where another gas-rich galaxy is nearby, its position is indicated with a green triangle and the size of the symbol relates to the gas mass of the subhalo (for aperture gas masses of $\log M_{\mathrm{Gas}} [\Msun]= 8,\,9,\,10,$ and $11$, see legend in the upper right corner of the panels in the right column). The fields-of-view of the plot in proper kpc is indicated in the bottom right of each figure and the database IDs for the host and TDGC are indicated in the title of each plot. }
		\label{fig:TDGC1}
	\end{center}
\end{figure*}

\begin{figure} 
	\begin{center}
		\includegraphics*[width=\linewidth]{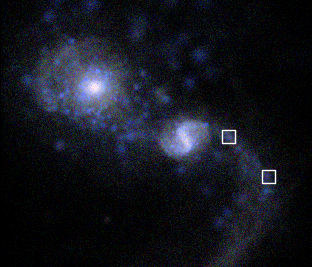}
		\caption{Composite gri colour image of the interacting galaxies 40017 and 1312231 (database GalaxyID), visualised with the radiative transfer code {\textsc {skirt}} (see text). The white squares indicate the location of TDGCs Ref-1 and Ref-2 within a faint tidal arm. The corresponding neutral gas column density distribution is shown in the bottom middle panel of Fig.~\ref{fig:TDGC1}. The extent of the figure is 170 pkpc.}
		\label{fig:TDGC1-skirt}
	\end{center}
\end{figure}

As TDGs originate from a more massive host galaxy, it is necessary to obtain information about their progenitors for a verification of a tidal origin of the TDG candidates. Merger trees that are set up to trace hierarchical structure formation cannot be used directly for the anti-hierarchical formation of TDGs, where lower mass galaxies originate from more massive ones. A galaxy that spawns a TDG will not have the TDG as its descendant as only a very small fraction of its original mass will end up in the TDG. We reverse the merger tree process for all TDG candidates that are selected in the database query by tracing back the 20 most bound gas particles of each TDG candidate to the previous snapshot. If a subhalo contains $> 50\%$ of the traced TDG particles, it is selected as candidate host galaxy of the TDG candidate. The potential host galaxy can then be traced forward to the selection snapshot by following the original merger tree.

We conclude that the TDG candidate consists of material from a more massive galaxy rather than from an in-falling gaseous substructure if (1) the host galaxy has a gas mass of $M_{\mathrm{host,gas}} > 10^9\Msun$ at both considered snapshots and (2) the traced TDG particles are on average closer to the centre of the host galaxy ($\bar{d}_{\mathrm{TDGC-host,tb}}$, where the suffix $\mathrm{tb}$ stands for ``traced back") than two gas half-mass radii or 70 pkpc to the progenitor of the host galaxy (inside the disk) and (3) the TDG at the selection snapshot is further away from the host galaxy than two gas half-mass radii: $d_{\mathrm{TDGC-host}}>2 \times R_{\mathrm{h,gas}}$ (outside the disk). As done above for $d_{\mathrm{TDGC-host}}$, an absolute maximum value of 70 pkpc is used for $\bar{d}_{\mathrm{TDGC-host,tb}}$ for cases where $R_{\mathrm{h,gas}}$ is not a good representation for the disk size. Increasing this value to 150 pkpc leads to two additional identifications that would be rejected as TDGCs, as the traced back particles are clearly far outside the gaseous disk on visual inspection.

Table~\ref{tab:parameter} summarizes the selection parameters described above. The parameters are as non-restrictive as possible while still assuring that the resulting TDGC can be marginally resolved with at least $\approx$ 44 gas particles (the median gas mass of the TDGC sample is resolved by more than 350 particles). For reference, EAGLE uses 58 neighbours for the SPH interpolation and 48 neighbouring particles to distribute stellar mass loss.

Limiting the host galaxy gas mass to $M_{\mathrm{host,gas}} > 10^9\Msun$ ensures that the TDGCs can still be identified if only 1$\%$ of the gas mass of the host galaxy collapses into the TDGC. Reducing the minimum host galaxy mass to $10^{8.5}\Msun$ results in a few additional identifications that do not satisfy the criteria for the traced-back particles. The final sample is therefore not sensitive to a reduction of  $M_{\mathrm{host,gas}}$. Increasing the minimum host galaxy gas mass to $10^{9.5}\Msun$, on the other hand, would remove two TDGCs that now are included in the final sample. As we are not aiming to quantify the TDGC population in EAGLE, the main result of this work is insensitive to the exact values of any of the selection parameters.

These selection criteria are not unique to TDGs: gravitationally bound objects that form in material that is ram pressure stripped from their host galaxies fall into the same category. This other type of dark matter free, anti-hierarchically formed objects was named ``fireballs" by \citet{yoshida_strange_2008} to distinguish them from TDGs. We therefore only further consider TDGs that originate from a host galaxy that is involved in a galaxy interaction. For the small number of TDGs identified here, an ongoing galaxy interaction is visually confirmed, while merger events are constructed from the database.

From the algorithmically determined sample, one candidate (Rej-1, Fig.~\ref{fig:TDGCRecal1}) was rejected based on its extremely low stellar metallicity ($\log_{10} Z/\Zsol < -3$, $> 1\sigma$ below the mass-metallicity relation). The stars in Rej-1 formed from very metal-poor gas very close to the snapshot where it was identified as a bound structure. TDGs inherit the enriched gas of their host galaxy but the star particles in Rej-1 have metallicities that are $\approx$ three orders of magnitude lower than the metallicity of their parent galaxy. 
A second candidate (Rej-2, Fig.~\ref{fig:TDGC5}) was rejected as its host galaxy is not involved in a galaxy interaction and morphologically it resembles a ``fireball" \citep[for a recent observation of star formation in ram pressure stripped gas, see e.g.][]{gullieuszik_gasp._2017}. For details of the rejected candidates, see Appendix~\ref{sec:rejected}.

\section{Results}\label{sec:results}

\begin{figure*} 
	\begin{center}
		\includegraphics*[height=5.8cm, trim={1cm 0.3cm 6.65cm 0.4cm},clip]{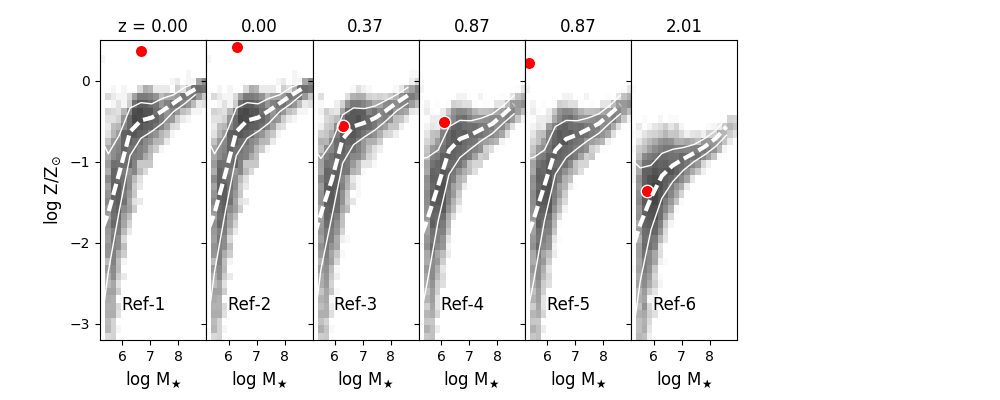}
		\includegraphics*[height=5.8cm, trim={2.55cm 0.3cm 12.cm 0.4cm},clip]{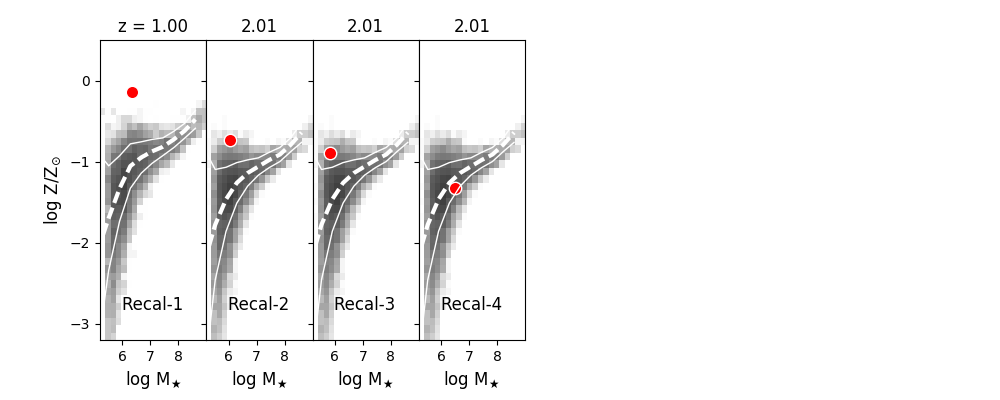}
		\caption{Stellar metallicity and masses for all TDG candidates (red points) together with the mass metallicity relation at the TDG selection redshift (greyscale histogram) for all subhalos with a DM mass of more than 10\% of their baryonic aperture mass and non-zero stellar metallicities. The histogram shows the logarithmic number of objects in each bin and the dashed white line is the median of the selected dwarf galaxy sample. The stellar metallicity is normalised to $\Zsol = 0.0127$. The left six panels are for TDG candidates identified in RefL0025N0752 and the right four panels are for TDG candidates identified in RecalL0025N0752.}
		\label{fig:MZrelation}
	\end{center}
\end{figure*}

Table~\ref{tab:TDGlist} lists the properties of the TDG candidates (6 from RefL0025N0752 and 4 from RecalL0025N0752) and their host galaxies both at the selection snapshot (SN) as well as at the previous snapshot (SN-1) identified between redshifts of 2 (snapshot 15) and 0 (snapshot 28). 

 The column density distributions of neutral hydrogen\footnote{The EAGLE particle data includes the hydrogen mass fraction for each SPH particle but the hydrogen mass is not further split up into its individual components (\ion{H}{i}, \ion{H}{ii}, H$_2$) as this would require computationally expensive radiative transfer and chemical network calculations. \citet{rahmati_evolution_2013} provide a fitting formula for the neutral (\ion{H}{i} + H$_2$) hydrogen fraction that is calibrated to smaller simulations with detailed radiation transport modelling. The approximation from \citet{rahmati_evolution_2013} includes self-shielding and depends on the density and temperature of the gas, as well as on the photoionisation rate. For consistency we use the redshift-dependent UV background from \citet{haardt_modelling_2001} as the same radiation field is used to calculate the cooling rates in EAGLE  \citep[from][]{wiersma_effect_2009}. Gas particles are limited by a pressure floor that is described as an equation of state. The gas temperature is set to $10^4$ K for particles with temperatures that are within 0.5 dex of the equation of state \citep[as in][]{bahe_distribution_2016}.} for the ten TDG candidates and their environments are presented in Fig.~\ref{fig:TDGC1} and Figs.~\ref{fig:TDGC2} to \ref{fig:TDGCRecal6}. More information about the individual TDGCs and their host galaxies is given in the caption of these figures.
 
Complementary to the gas distributions, Fig.~\ref{fig:TDGC1-skirt} shows the interacting pair of galaxies that form the two TDGCs Ref-1 and Ref-2 in optical / near-infrared colours.  Mock rgb images in the SDSS gri bands are generated by post-processing the stars and gas with the Monte Carlo radiative transfer code {\textsc{skirt}} \citep{baes_radiative_2003, baes_efficient_2011, camps_skirt:_2015}. This code allows for the accurate computation of the absorption and scattering of stellar light off of dust grains, taking into account the full 3D structure of the stars and gas in galaxies. We adopt the same radiative transfer parameters as \citet{trayford_optical_2017}, but note that we do not perform their subsampling of young stellar component or modelling of \ion{H}{II} regions, but rather model all stellar particles using {\textsc{galexev}} \citep{bruzual_stellar_2003} SEDs, assuming a \citet{chabrier_galactic_2003} initial mass function. Images are obtained at 0.477 micron, 0.6231 micron, and 0.7625 micron (the effective wavelengths of the SDSS g, r, and i filters, respectively), and reprocessed into RGB images following the procedure of \citet{lupton_preparing_2004}. Despite the low surface brightness of the tidal arm, its blue colour is indicative of a young stellar component. The $z=0$ star formation rates are $0.2$ (Ref-2) and $0.3\,\Msun \,\mathrm{yr}^{-1}$ (Ref-1) and the youngest star particles in both TDGCs formed less than 50 Myr before the snapshot.
 
Observed TDGs have metallicities that are elevated with respect to the mass-metallicity relation of hierarchically formed dwarf galaxies (DGs) as their gas is already pre-enriched in the host galaxy 
\citep[see e.g.][]{duc_formation_2000, weilbacher_tidal_2003,croxall_chemical_2009}. In Fig.~\ref{fig:MZrelation} the stellar metallicity of the TDG candidate is compared to the stellar metallicities of DGs with similar stellar masses. Seven out of ten TDGCs have elevated metallicities for their stellar mass ($>$1$\sigma$ above the average) and three TDGCs have average metallicities (within $\pm$1$\sigma$). Note that the mass-metallicity relation is redshift-dependent and the TDGCs that have elevated metallicities at $z=2$ would lie on the $z=0$ relation (Fig.~\ref{fig:MZrelation}) if they evolved passively until the present-day, as expected from chemical models \citep{recchi_mass-metallicity_2015}.

The TDGCs selected in our sample have gas masses from $4.8\times10^7\Msun$ to $5.2\times10^8\Msun$ (median: $8.4\times 10^7\Msun$) and stellar masses of $\approx 10^6\Msun$ (median: $1.7 \times 10^6 \Msun$). At their selection snapshot they are at a median distance of 38 pkpc from the centre of their host galaxy. Two TDGCs at redshift 0 (Ref-1, Ref-2) and two TDGCs at redshift 2 (Recal-2 and Recal-3) are formed out of the same host galaxy. This could be a hint that the TDG formation per galaxy interaction is more efficient than what would be derived from the total numbers here, but the time window during which they can be identified in the simulation might be short relative to the timespan between two snapshots. This is not necessarily related to the physical survival time of TDGCs, but could be related to the poorly sampled feedback or the pressure floor that prohibits the TDGs from collapsing further, increasing their binding energy (see Sec~\ref{sec:discussion}).

All but one TDGC in the reference simulation are identified after $z=1$ while all TDGCs in the re-calibrated simulation formed prior to $z=1$. Only more detailed simulations can show if this behaviour is indeed linked to the different feedback descriptions or if it is a spurious result caused by the small total number of TDGs that can be investigated here due to the limitations discussed in Sec.~\ref{sec:discussion}.

The sample of observed TDGs is not limited to binary mergers, their formation is also observed in denser environments, such as compact groups of galaxies. For examples see e.g. \citet{temporin_candidate_2003} for a TDG in CG J1720-67.8, \citet{torres-flores_star_2009, torres-flores_star-forming_2014} in Robert's Quartet, \citet{mendes_de_oliveira_candidate_2001} in Stephan's Quintet, \citet{eigenthaler_star_2015} in HCG 91, and \citet{de_mello_searching_2008} for two candiates in HCG 100. The high space density in combination with the low velocity dispersion of the galaxies in compact groups is a favourable environment for the formation of extended tidal arms compared to both field galaxies and galaxies in clusters. 
In spite of the small sample size one TDGC presented here (Ref-5, Fig.~\ref{fig:TDGC7}) forms in a compact group of galaxies, and another TDGC (Ref-3, Fig.~\ref{fig:TDGC4}) forms in a high speed encounter with a relative velocity of $\approx 430\kms$\footnote{Observations show that TDGs can indeed form in high-speed encounters. As an example, the TDGs in the tidal debris around NGC5291 likely formed in a galaxy collision with a relative velocity at impact of $1250\kms$, according to numerical models of \citet{bournaud_missing_2007}.}. While the wide range of different environments that result in the formation of TDGCs in EAGLE is promising, the low number of candidates does not allow for a quantitative investigation on the impact of the environment on the formation of TDGs.

The higher gas fractions of high redshift ($z \approx 2$) discs trigger gravitational instabilities that can lead to the formation of massive gas clumps. In galaxy interactions these clumps can be expelled from their parent galaxy \citep[see e.g.][]{bournaud_hydrodynamics_2011, fensch_high-redshift_2017}. This is potentially a separate formation scenario for TDGs compared to the gravitational collapse of gas directly in the tidal arm. TDGCs in EAGLE indeed form in decreasingly prominent tidal arms with increasing redshift and the host galaxies of the $z=2$ TDG candidates (Ref-6, Recal-2, Recal-3) clearly show a clumpy morphology. {\textsc{subfind}} does not identify the TDG progenitor in the previous snapshot at $z=2.24$, but it remains unclear if this is a clear indication that the gravitational collapse occurs outside the galaxy disc, or if this can be attributed to the lack of resolution. Higher resolution simulations can help to quantify the difference between low and high-z TDG formation. 

Cosmological simulations provide an unbiased view on the dependence of the TDG formation rate on both redshift and environment. The exemplary TDGCs presented in this work cover a wide range of possible formation scenarios and show that state-of-the-art cosmological simulations are approaching the necessary resolution to study these questions quantitatively. 

\section{Discussion}\label{sec:discussion}

\begin{table*}
\centering
\caption{List of identified TDG candidates. The last two rows list the median and mean values for all TDGCs. Column 1:  consecutive identification number for the TDG candidates within this work; column 2:  snapshot number where the TDGC is identified; column 3: redshift; columns 4-6: TDG candidate properties: database identifier (column 4), aperture gas mass (column 5), and aperture stellar mass (column 6); columns 7 - 11: host galaxy properties at the snapshot where the candidate is selected: database identifier (column 7), distance to the centre of potential of the host galaxy (column 8), aperture gas mass (column 9), aperture stellar mass (column 10), and aperture dark matter mass (column 11); columns 12 - 15: as columns 8 - 11 but for the snapshot before the TDGC selection. The distance (column 13) here is the mean distance between the traced-back TDG particles and the progenitor of the host galaxy. }\label{tab:TDGlist}
	\begin{tabular}{lrrrrrrrrrrrrrrrr}
	\#&SN	& z 	& \multicolumn{3}{c}{TDG candidate} & \multicolumn{5}{c}{Host galaxy candidate $[$SN$]$} &	\multicolumn{4}{c}{Host galaxy candidate $[$SN-1$]$} 						\\
		&&	&ID	& $M_{\mathrm{Gas}}$/	&$M_{\mathrm{Stars}}$/	&ID		&d/		&$M_{\mathrm{Gas}}$/	&$M_{\mathrm{Star}}$/	&$M_{\mathrm{DM}}$/	&$\bar{d}$/&$M_{\mathrm{Gas}}$/ &$M_{\mathrm{Star}}$/	&$M_{\mathrm{DM}}$/		 \\
		&&	&	&$10^6\Msun$		&$10^6\Msun$		&		&pkpc	&$10^9\Msun$		&$10^9\Msun$			&$10^9\Msun$			&pkpc		&	$10^9\Msun$	&$10^9\Msun$			&$10^9\Msun$				\\
\hline
 \multicolumn{15}{l}{RefL0025N0752}\\
Ref-1       &28     &0.0    &7592323        &60.1   &4.6    &40017  &24.0   &3.4    &14.8   &47.3   &15.5   &10.2   &19.1   &121.7\\
Ref-2       &28     &0.0    &7592324        &47.9   &2.1    &40017  &50.9   &3.4    &14.8   &47.3   &17.4   &10.2   &19.1   &121.7\\
Ref-3       &24     &0.4    &1326560        &72.5   &2.0    &45588  &26.8   &2.8    &7.3    &26.8   &41.2   &7.3    &8.5    &115.7\\
Ref-4       &20     &0.9    &785066 	      &92.6   &1.3    &785838 &10.1   &1.3    &0.4    &4.8    &13.0   &3.9    &0.3    &22.8\\
Ref-5       &20     &0.9    &6776662        &386.6  &0.2    &1544074        &73.8   &6.0    &3.6    &74.0   &52.0   &6.1    &3.0    &76.4\\
Ref-6       &15     &2.0    &1858256        &107.0  &0.6    &1861291        &21.8   &19.1   &11.8   &189.5  &51.4   &15.1   &7.5    &181.4\\
\hline
 \multicolumn{15}{l}{RecalL0025N0752}\\
Recal-1       &19     &1.0    &1837214        &63.2   &2.2    &6712351        &47.8   &6.5    &9.7    &104.5  &21.7   &14.0   &8.2    &133.8\\
Recal-2       &15     &2.0    &1819619        &137.9  &1.1    &1822350        &41.0   &17.0   &10.9   &163.3  &42.4   &18.9   &7.0    &192.3\\
Recal-3       &15     &2.0    &1819751        &75.4   &0.7    &1822350        &34.5   &17.0   &10.9   &163.3  &39.4   &18.9   &7.0    &192.3\\
Recal-4       &15     &2.0    &1655441        &519.9  &3.3    &6836598        &45.7   &3.7    &0.2    &35.1   &14.8   &7.2    &0.1    &51.2\\
			\hline
\multicolumn{2}{l}{median} & 0.95 	&			&84.0	 &1.7	&			&37.8 & 4.9 & 10.3 & 60.7 & 30.6 & 10.2 & 7.3 	&121.7 \\
			\hline
\multicolumn{2}{l}{mean}& 1.12		&			&156.3	&1.8	&			&37.6 &8.0& 8.4 &85.6 &30.9 &11.2 & 8.0 & 120.9 \\	
	\end{tabular}
\end{table*}

The initial mass of a baryonic particle in the high-resolution EAGLE simulations is $m_{\mathrm{g,star}} = 2.26 \times 10^5\,\Msun$ (the individual particle mass can vary due to stellar mass loss) and the median gas mass of the identified TDGCs is $8.4 \times 10^7 \Msun$ (median stellar mass: $1.7\times10^6\Msun$). The number of gas particles ($\approx 370$ for the median gas mass) is therefore sufficient to study the formation of these gravitationally-bound structures as a result of galaxy interactions, but for the subsequent evolution of TDGs, detailed modelling of their internal stellar feedback processes, in addition to the external tidal field, is crucial. Star formation in EAGLE is implemented stochastically because of  the fixed mass resolution of stellar particles. For TDGs and other low-mass objects the stellar population is not fully sampled. For the initial TDGC selection the stellar mass is required to be non-zero as the stellar metallicity of TDGCs is compared to the stellar mass-metallicity relation. As the total stellar mass of each TDG candidate is close to the particle mass and star formation is implemented stochastically, the selection on non-zero stellar masses is partly determined by the stochastic implementation of the star formation process which introduces a scatter in the identification. If, on the other hand, a TDGC stochastically forms too many stars, the effects of stellar feedback in both the energy budget and the metal enrichment are likely overestimated. Future higher-resolution simulations (cosmological boxes or zoom-in simulations) will allow for a less stochastic TDG identification.

Extended tidal arms in interacting galaxies are typically observed in \ion{H}{i}, but the neutral gas phase is not yet accurately modelled in cosmological simulations. In order to prevent artificial fragmentation, a pressure floor is imposed that keeps the gas (dynamically) warm at high densities. Dense gas in EAGLE follows an equation of state and the neutral gas phase (and the star formation within it) is described by sub-grid recipes. For tidal arms that consist mainly of neutral gas, the artificial pressure floor can prevent the collapse into small structures such as TDGs. Increasing the resolution would enable the pressure floor to be lowered and in combination with a more detailed treatment of the neutral gas phase, the collapse of structures within tidal arms could be investigated more quantitatively. 

As another side effect of the pressure floor, the spatial extent of the TDGCs is not very reliable and therefore not investigated here. Especially for the high-redshift candidates, the dark matter-free bound structures could also be the birthplaces of massive compact star clusters that are observed as globular clusters in the local Universe, rather than extended high redshift TDGs. The minimum gravitational softening length of 350 pkpc precludes the discrimination between a compact massive star cluster with a half-mass radius of few ppc that formed in one single star formation event (e.g. globular cluster progenitor) and a more extended object with more moderate star formation.

To identify TDGs that form within a cosmological simulation and to study the properties of their host galaxies and the interactions they are involved in, the halo-finding algorithm has to accurately assign particles to substructures even during close encounters. \citet{behroozi_major_2015} compared five halo-finding algorithms during major mergers and found significant differences in the substructures they identified. In one example they started with two identical haloes at rest at a separation of 2700 kpc. During the encounter, the mass ratio between the two substructures that {\textsc{subfind}} identifies is much lower than 1 (see their figure 4) and during the late stages of the merger one of the substructures disappears completely for two snapshots and reappears again afterwards. This interaction would not be classified as a major merger as the mass ratio from {\textsc{subfind}} is lower than 1:4 for most of its evolution. For the construction of the merger trees for the EAGLE simulation, the disappearance and reappearance of subhalos in {\textsc{subfind}} is taken into account \citep{qu_chronicle_2017}. However, algorithms that track particles in time such as {\textsc{hbt}} \citep[Hierarchical Bound-Tracing algorithm,][]{han_resolving_2012} or that identify structures in phase-space such as the {\textsc{rockstar}} halo finder \citep{behroozi_rockstar_2013} or {\textsc{velociraptor}}  \citep{elahi_peaks_2011} conserve the correct mass ratio better in these situations.

\begin{figure*}
	\begin{center}
		\includegraphics*[width=\linewidth,trim={1.2cm 0.3cm 1.cm 0cm},clip]{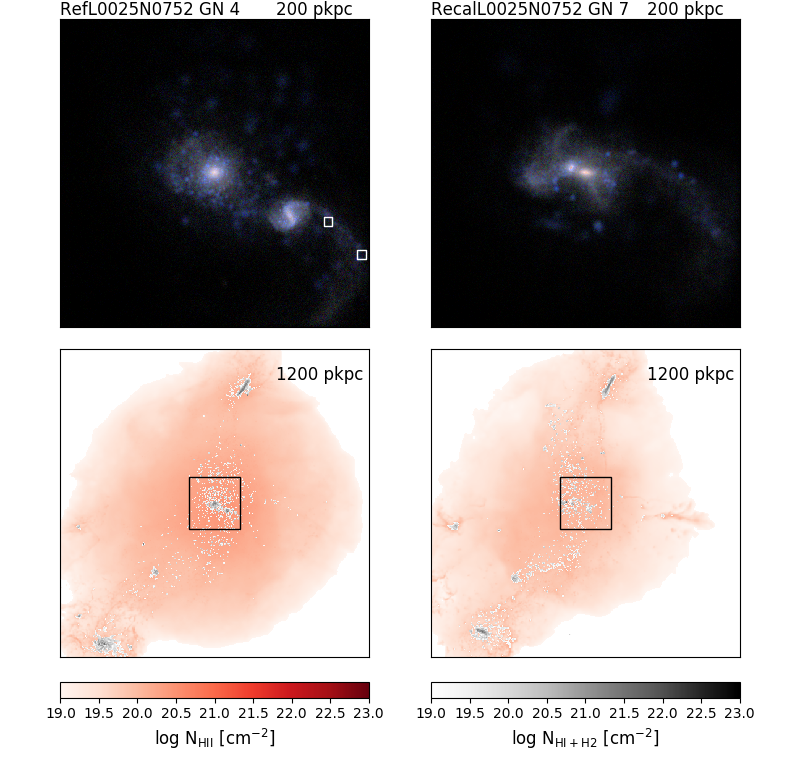}
		\caption{The small-scale (200 pkpc, top panels) and large-scale (1200 pkpc, bottom panels) environments at the same simulation box coordinates in RefL0025N0752 (left panels) and RecalL0025N0752 (right panels) at $z=0$. The bottom panels show the column density of ionised hydrogen (\ion{H}{ii}; red) and neutral hydrogen (\ion{H}{i} and H$_2$; grey) that belongs to the FoF group (Nr. 4 in Ref and Nr. 7 in Recal, see text). The neutral hydrogen fraction is calculated with the fitting formula from \citet{rahmati_evolution_2013}. The top panels show the composite gri colour image (as Fig.~\ref{fig:TDGC1-skirt}) of the region indicated with a black square. The interacting galaxy pair highlighted in the top left panel includes the host galaxy of TDGCs Ref-1 and Ref-2 (white squares). While both simulations have an interacting pair of galaxies at the same position as well as a similar large-scale environment, the participation galaxies have different gas fractions and are at a slightly different merger stage. }
		\label{fig:Environment}
	\end{center}
\end{figure*}

The EAGLE database lists the sizes of the stellar disks of galaxies in terms of their half-mass radii for different aperture sizes but for the gaseous disk only the half mass radius of the complete subhalo is included. Especially for subhalo 0, the background halo to which all particles are assigned that are bound to the FoF group but not to any other substructure, the half-mass radius can be very large (several hundred kpc). This size does not reflect the properties of the gas distribution of the central galaxy because it includes the low-density gas in the circumgalactic medium. Determining whether a substructure is located inside or outside the gaseous disk is therefore not straightforward. Including the half-mass radius for gas within an aperture in the database, as is done for the stellar half-mass radius, would already partially solve this problem. For a quantitive study this could be calculated for different aperture sizes analogously to the stellar half-mass radii for the full galaxy population in EAGLE. For the identification of the exemplary objects presented here, we imposed a limit on the absolute distance in proper kpc to account for central galaxies with too large half-mass radii (see Sec.~\ref{sec:stepselection}). 

The merger trees for cosmological simulations in general exclusively trace hierarchical structure formation.  Future simulations with higher resolution can resolve  the birthplaces of anti-hierarchically formed structures, such as globular clusters, TDGs or ram-pressure stripped dwarf galaxies. These objects could be more easily identified and studied in greater detail if a ``reversed" merger tree, in addition to the information about the hierarchical growth of structures, is constructed for all bound structures in the box.

For the purposes of this study, the temporal spacing of the snapshots in the EAGLE database is coarse. For objects whose formation, but not yet evolution, can be traced, the long duration between snapshots is challenging and could result in a large underestimation of formation rates in more quantitative studies. Only for the high-redshift object detected here, the TDGC is already starting to form at the snapshot prior to the selection snapshot, while for the low-redshift objects the traced-back TDGC particles are still distributed over a large volume within the host galaxy. A higher snapshot frequency would also enable better constraint of the geometry of the galaxy interaction. For RecalL0025N0752 additional outputs with finer time resolution (``snipshots") exist but not for RefL0025N0752. As in addition the database can be queried only for snapshot data, we do not include the snipshot data for RecalL0025N0752.

The two simulation runs considered have the same initial conditions and only vary in terms of the feedback parameters. In this work, they are used as two independent simulations to increase the expected low number of TDGCs. The difference between the galaxy populations between the simulations with different subgrid parameters in EAGLE is explained in \citet{schaye_eagle_2015} and extensively studied in \cite{crain_eagle_2015}. Fig.~\ref{fig:Environment} illustrates the variation for the FoF halo that contains TDGCs Ref-1 and Ref-2. For this comparison the database was queried for the FoF group in Recal that is closest in location as well as in logarithmic group mass to its analogue in Ref. The large-scale column density distributions for FoF group 4 (Ref; $M_{\mathrm{Crit200}} = 4.11 \times 10^{12}\,\Msun$) and FoF group 7 (Recal;  $M_{\mathrm{Crit200}} = 3.90 \times 10^{12}\,\Msun$) are similar, as expected. 

The neutral hydrogen distribution within the FoF group varies significantly because of the different feedback parameter values whose effects accumulate as the simulation progresses. In the centre of both FoF groups two galaxies are in a close encounter but the stage of the merger as well as the gas and stellar masses of the involved galaxies differ (see Table.~\ref{tab:RefRecal} for details). The total gas mass within a 30 pkpc aperture of the interacting galaxy pair is a factor of 2.5 larger and the distance between the galaxies is a factor of 6 larger in Ref compared to the galaxy pair in Recal. This explains why two distinct gas disks are visible in Ref (see the inset of the left panel of Fig.~\ref{fig:Environment}) while the gas morphology is very perturbed without any clear gas disks in Recal (see the inset of the right panel of Fig.~\ref{fig:Environment}). We therefore do not expect to identify the same TDGC in the two simulation boxes, despite their identical initial conditions. For future simulations, where the formation of TDGs is better resolved, the influence of stellar feedback on the evolution of TDGs can be studied, but this is not yet possible and therefore the TDG candidates in the simulations are studied independently.

\begin{table}
\caption{Properties of the interacting pair of galaxies in the centre of FoF groups nr. 4 (Ref) and 7 (Recal) that are illustrated in the inset of Fig.~\ref{fig:Environment}. The more massive galaxy in each pair is denoted ``Gal 1" and the less massive galaxy ``Gal 2". For all four galaxies the IDs from the public database and their stellar and gas masses within a 30 kpc aperture are listed. Finally, the distance in proper kpc between each galaxy pair at $z=0$ is given in the last row. Note that the galaxy with GalaxyID = 40017 is the host galaxy of TDGCs Ref-1 and Ref-2 (Figs.~\ref{fig:TDGC1} and \ref{fig:TDGC2}). }
\begin{center}
\label{tab:RefRecal}
\begin{tabular}{lrrrr}
\hline
 & \multicolumn{2}{c}{Ref} & \multicolumn{2}{c}{Recal} \\
 &Gal 1 &Gal 2& Gal 1 & Gal 2 \\
\hline
GalaxyID 		      &  1312231 &40017 & 1378064 & 6774613 \\
$M_{\mathrm{Gas}}[10^9 \Msun]$&  7.9 &  3.4& 3.2 &1.4\\
$M_{\mathrm{Star}}[10^9 \Msun]$&  38.2 &14.8  &32.1 &5.7\\
d $[$pkpc$]$	&	\multicolumn{2}{c}{61} & \multicolumn{2}{c}{10}  \\
\hline
\end{tabular}
\end{center}
\end{table}

\section{Conclusion}\label{sec:conclusions}

In spite of the limitations in box size, number of snapshots (low number of mergers), resolution, as well as in the halo finding algorithm, we identified for the first time TDGCs that form self-consistently in a cosmological simulation. Therefore, the distribution of parameter describing the orbits (e.g. eccentricity, impact parameter) and the individual galaxies (e.g. mass, size, angular momentum, gas fraction) in galaxy encounters within cosmological simulations is set by cosmic structure formation as well as the mass assembly and star formation history of the participating galaxies. 

We use the public database to identify TDGCs in the two high-resolution boxes (RefL0025N0752 and RecalL0025N0752) of the EAGLE suite of cosmological simulations and select ten of them through a combination of their database properties as well as inspection of the host galaxies and their immediate surroundings at the two consecutive snapshots that bracket the time when they become gravitationally bound structures. The TDGCs in EAGLE have median gas masses of $8 \times 10^7\Msun$ and contain young stars with elevated metallicities relative to the mass-metallicity relation. They originate from host galaxies with median gas masses of $10^{10}\Msun$ ($5\times10^9\Msun$) and gas fractions of 60\% (32\%) at the snapshot before (at) the formation of the TDGC.

In addition to the identified TDGCs, one rejected candidate (Rej-2) resembles ram pressure stripped gas that is gravitationally bound and star forming (``fireballs"), which, whilst not the focus of this study, is clearly an interesting class of object.

We have shown that it is already possible to find TDGCs and similar dark matter free, bound objects in cosmological simulations and we conclude that with the improvements discussed in Sec.~\ref{sec:discussion}, especially in terms of resolution and halo finders, the next generation of cosmological simulations will allow a more quantitative investigation of the formation of TDGs in a $\Lambda$CDM Universe.

\section*{Acknowledgements}
We would like to thank the referee, Frederic Bournaud, for a careful review of the manuscript. The valuable comments and suggestions helped to improve the clarity of the publication. SP and CB acknowledge support from the European Research Council under the European Union's Seventh Framework Programme (FP7/2007- 2013)/ERC Grant agreement 278594-GasAroundGalaxies. RAC is a Royal Society University Research Fellow. We acknowledge the Virgo Consortium for making their simulation data available. The EAGLE simulations were performed using the DiRAC-2 facility at Durham, managed by the ICC, and the PRACE facility Curie based in France at TGCC, CEA, Bruy\`eres-le-Ch\^atel. The analysis in this work used the DiRAC Data Centric system at Durham University, operated by the Institute for Computational Cosmology on behalf of the STFC DiRAC HPC Facility (www.dirac.ac.uk). This equipment was funded by BIS National E-infrastructure capital grant ST/K00042X/1, STFC capital grants ST/H008519/1 and ST/K00087X/1, STFC DiRAC Operations grant ST/K003267/1 and Durham University. DiRAC is part of the National E-Infrastructure.



\bibliographystyle{mnras}
\bibliography{TDGs} 




\appendix

\section{Visual impression for all TDGCs} \label{sec:TDGfigures}

\begin{figure*}
	\begin{center}
		\includegraphics*[width=0.75\linewidth]{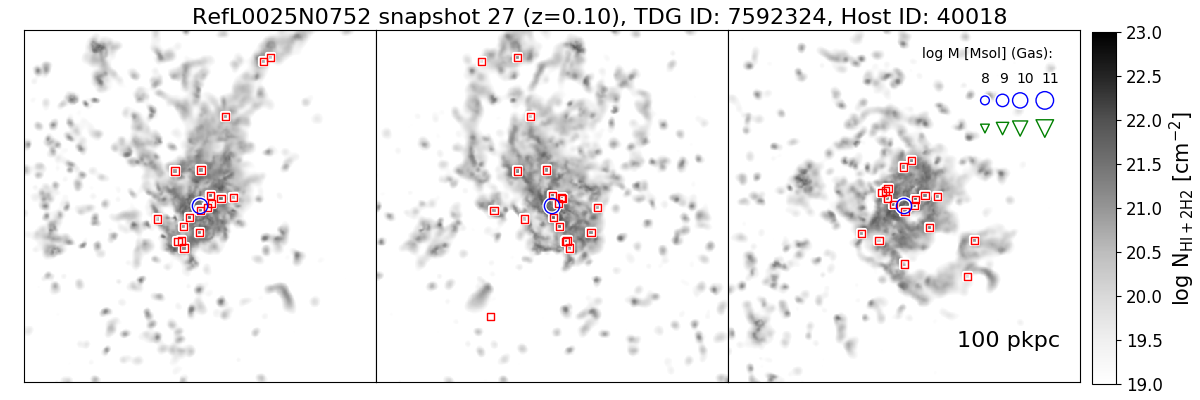}
		\includegraphics*[width=0.75\linewidth]{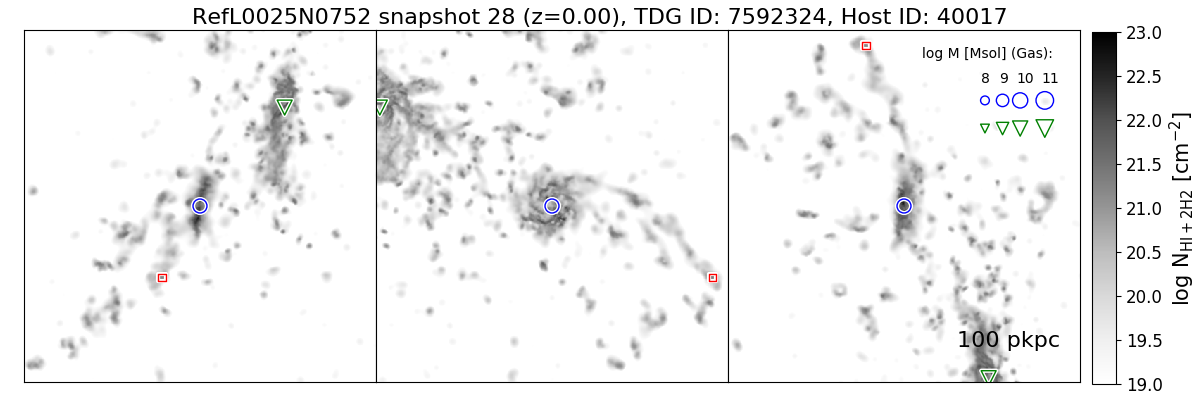}				
		\caption{As Fig.~\ref{fig:TDGC1} but for Ref-2. TDGCs Ref-1 and  Ref-2 are both identified at $z=0$ and originate from disk material of the same host galaxy. Their stellar metallicity is one order of magnitude higher than is typical for other dwarf galaxies of the same mass. They are both located in a clear tidal arm. The baryonic mass ratio of the two interacting galaxies is 2.5 at the selection time (bottom panel).}
		\label{fig:TDGC2}
	\end{center}
\end{figure*}

\begin{figure*} 
	\begin{center}
		\includegraphics*[width=0.75\linewidth]{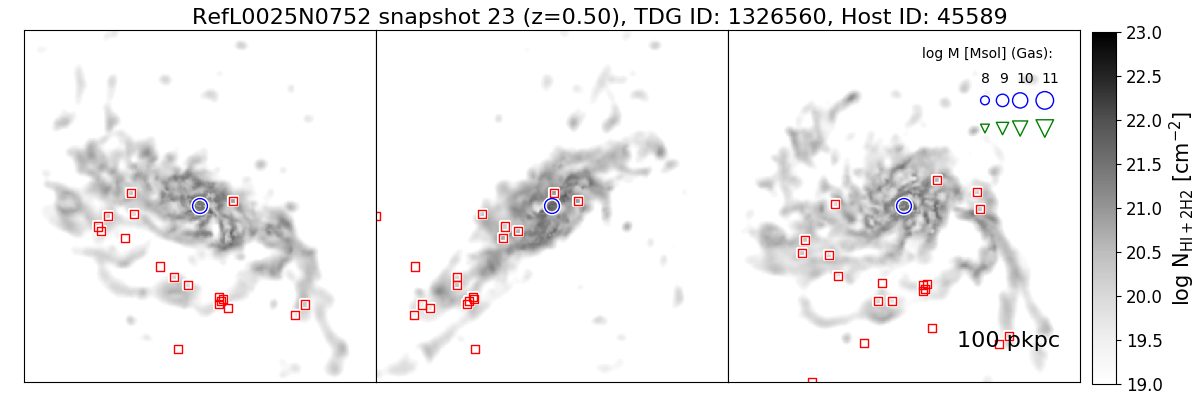}
		\includegraphics*[width=0.75\linewidth]{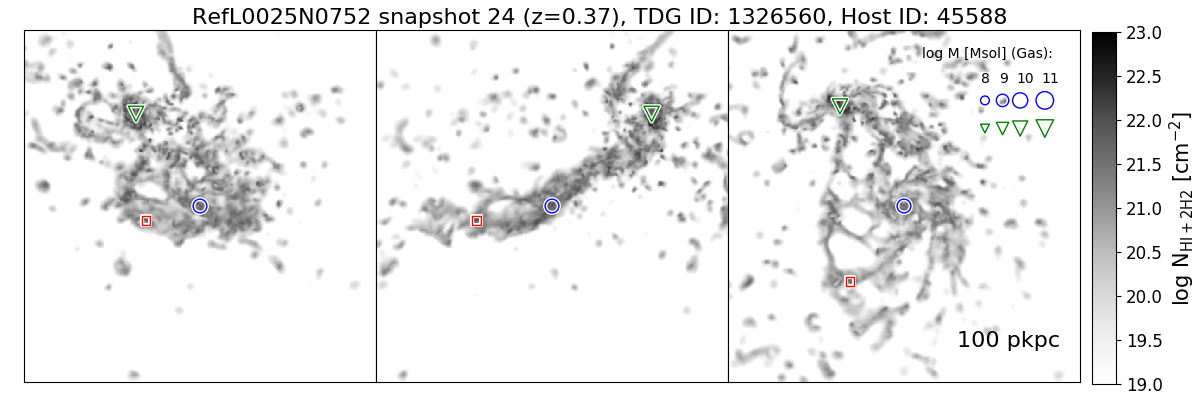}			
		\caption{As Fig.~\ref{fig:TDGC1} but for Ref-3. The traced-back particles are clearly located in the disk of their host galaxy at the snapshot before the TDG selection. At the next snapshot, the host galaxy is interacting with the central galaxy of the same FoF group. At the selection snapshot the two gas-rich galaxies are in a close high-speed encounter with a relative velocity of $\approx 430\kms$ and do not merge until redshift 0 (4 Gyr later). The stellar metallicity of the TDGC lies close to the mean mass-metallicity relation of hierarchically formed dwarf galaxies (see Fig.~\ref{fig:MZrelation}).}
		\label{fig:TDGC4}
	\end{center}
\end{figure*}

\begin{figure*} 
	\begin{center}
		\includegraphics*[width=0.75\linewidth]{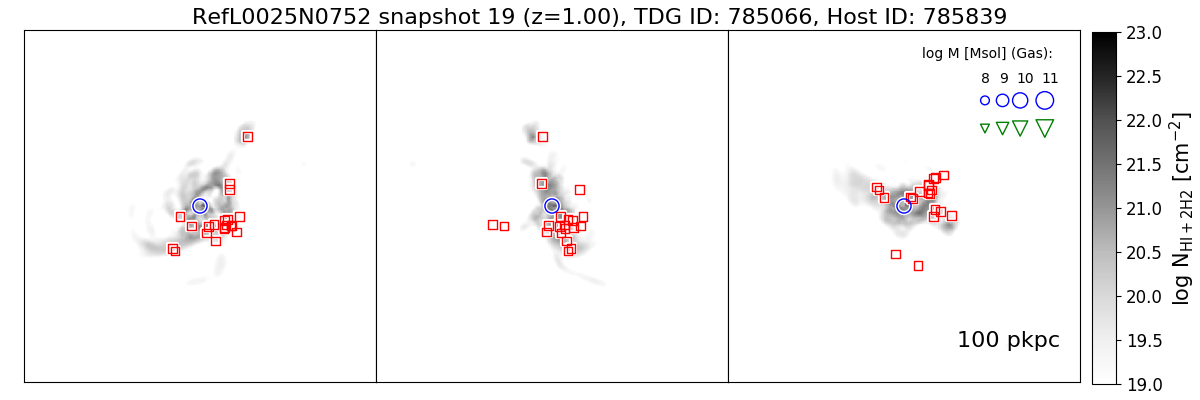}
		\includegraphics*[width=0.75\linewidth]{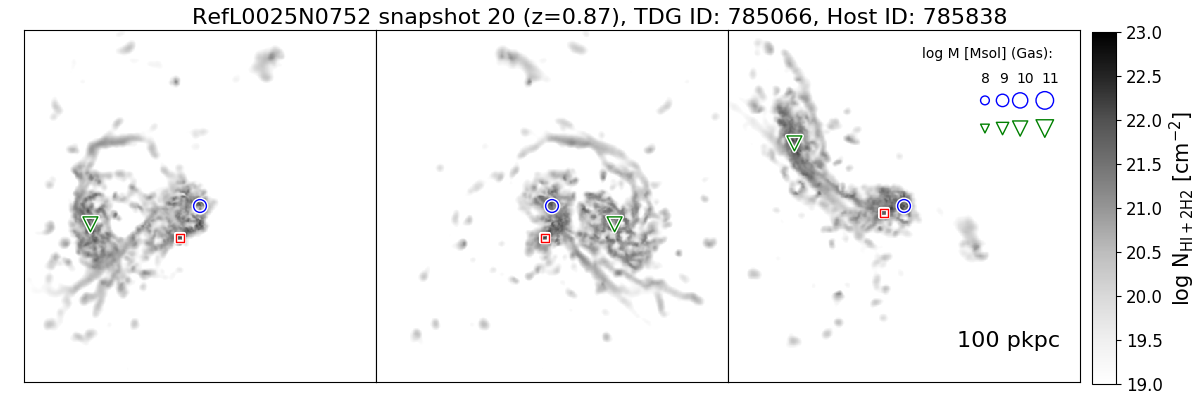}
		\caption{As Fig.~\ref{fig:TDGC1} but for Ref-4. The traced back particles are located within the gaseous disk of their host galaxy and the host galaxy is involved in a close galaxy interaction with a baryonic mass ratio of 1:13  (gas mass ratio 1:6) at the TDGC selection redshift. For these low mass ratios it is still possible to produce long tidal arms if the host galaxy is the less massive one as in this case. At the selection snapshot the TDGC is very close to its host galaxy. Tracing the TDG particles forward to the next snapshot reveals that they did not fall back onto the host galaxy but are being ejected to an average distance of $\approx$ 100 pkpc. Due to the limited resolution, it is not possible to determine for how long this TDGC would stay gravitationally bound. The stellar metallicity of Ref-4 is slightly enhanced relative to the mass-metallicity relation (see Fig.~\ref{fig:MZrelation}). }
		\label{fig:TDGC6}
	\end{center}
\end{figure*}

\begin{figure*} 
	\begin{center}
		\includegraphics*[width=0.75\linewidth]{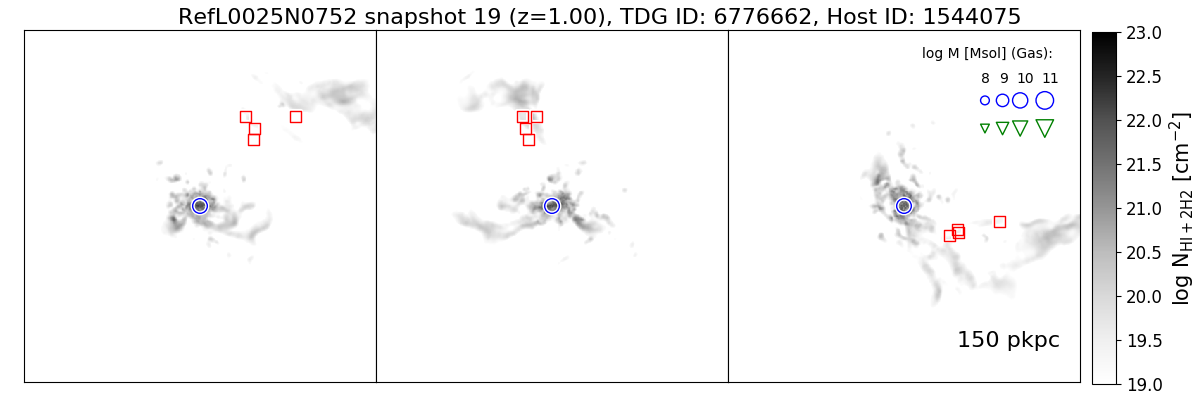}
		\includegraphics*[width=0.75\linewidth]{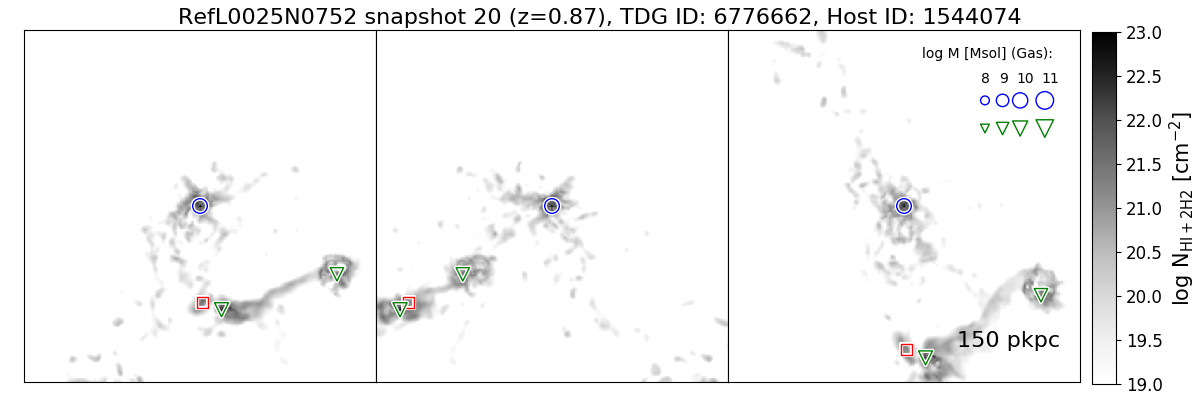}
		\caption{As Fig.~\ref{fig:TDGC1} but for Ref-5. This TDGC is part of a complicated galaxy interaction involving four galaxies (only three are shown here). The traced back particles are not clearly located in the disk at the snapshot before the TDG selection as the disk is already very perturbed. The host galaxy merges until the next snapshot ($z=0.74$) with the 2.7 times less massive galaxy (baryonic mass ratio) closest to the TDGC in the bottom panels. The stellar metallicity is clearly above the mass-metallicity relation. This example illustrates that the formation of TDGCs is not necessarily limited to galaxy pairs as studied in idealised galaxy interaction simulations. }
		\label{fig:TDGC7}
	\end{center}
\end{figure*}

\begin{figure*} 
	\begin{center}
		\includegraphics*[width=0.75\linewidth]{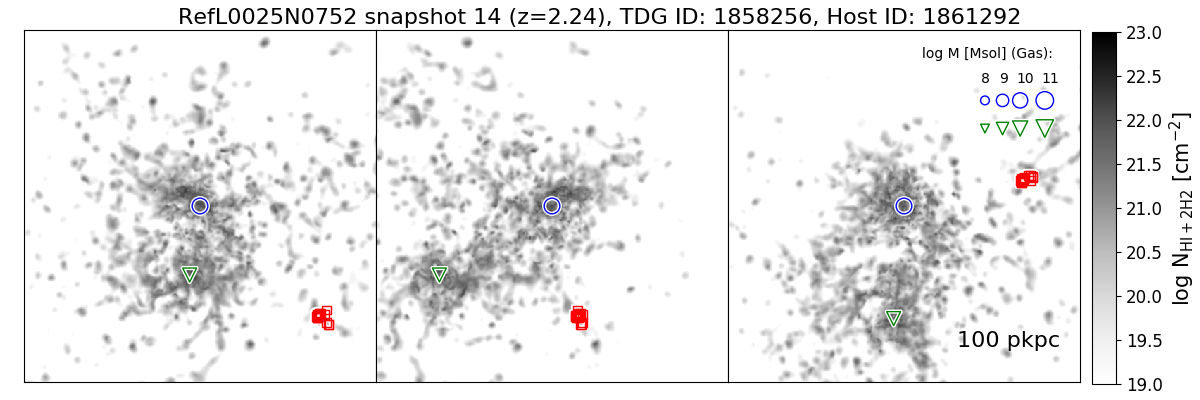}
		\includegraphics*[width=0.75\linewidth]{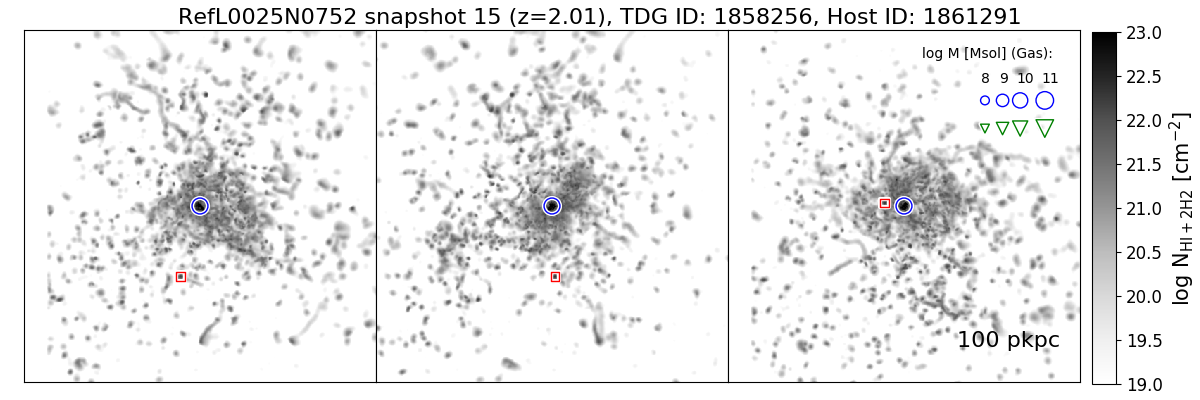}	
		\caption{As Fig.~\ref{fig:TDGC1} but for Ref-6. The host galaxy merges with another galaxy with 5 times less baryonic mass before the TDG selection snapshot. The traced-back particles are already outside the disks of the merging galaxies but as the difference in lookback time between snapshots 14 and 15 is only around 330 Myr, the traced back TDGs particles are already located close to each other and in the process of collapsing into the TDGCs seen at snapshot 15. This is supported by tracing back the particles for another 300 Myr to snapshot 13, where the particles are already within one clump, but {\textsc{subfind}} only identifies the clump as gravitationally bound at snapshot 15. The stellar metallicity follows the mass-metallicity relation.}
		\label{fig:TDGC9}
	\end{center}
\end{figure*}


\begin{figure*} 
	\begin{center}
		\includegraphics*[width=0.75\linewidth]{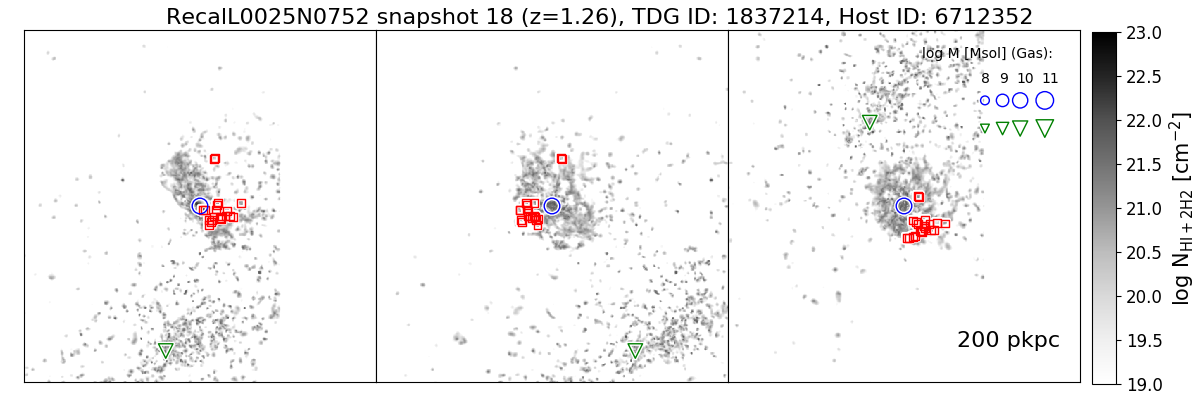}
		\includegraphics*[width=0.75\linewidth]{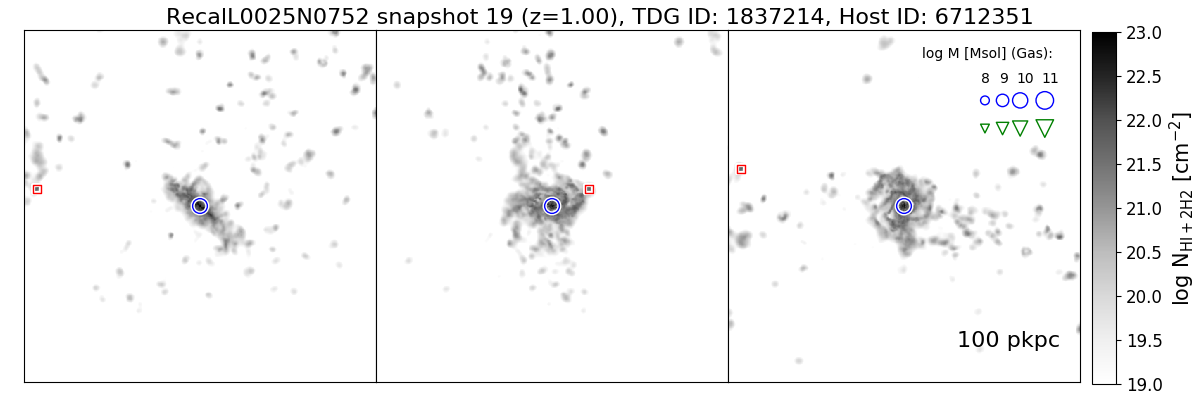}
		\caption{As Fig.~\ref{fig:TDGC1} but for Recal-1. As seen in Fig.~\ref{fig:MZrelation}, Recal-1 shows a very clear excess stellar metallicity. The traced back particles are located within the disk of the host galaxy and an interaction is visible in the snapshot before the TDG selection (top panel). The other involved galaxy (triangle symbol) is twice as massive in baryons as the TDGC host galaxy but has a low gas fraction ($\approx 11 \%$ of the baryonic mass in the 30 pkpc aperture). The interaction does result in a merger until $z=0$.}
		\label{fig:TDGCRecal2}
	\end{center}
\end{figure*}

\begin{figure*} 
	\begin{center}
		\includegraphics*[width=0.75\linewidth]{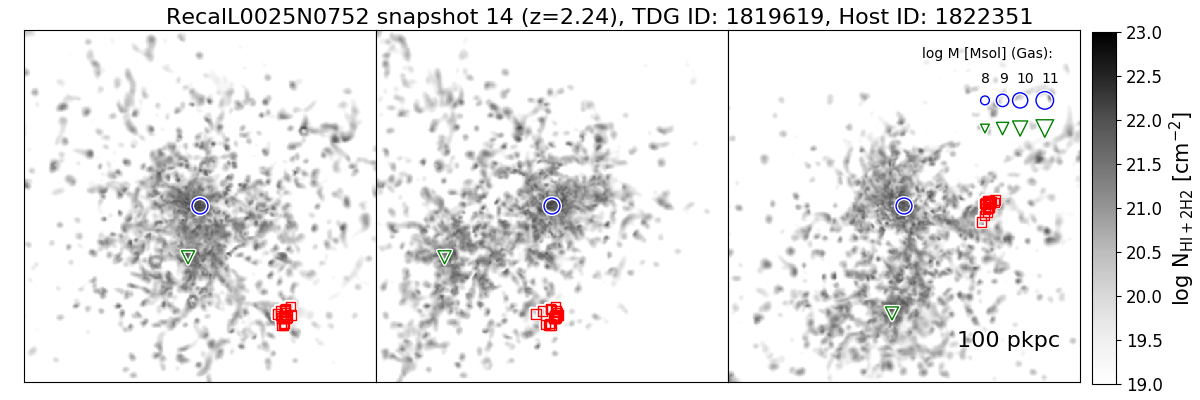}
		\includegraphics*[width=0.75\linewidth]{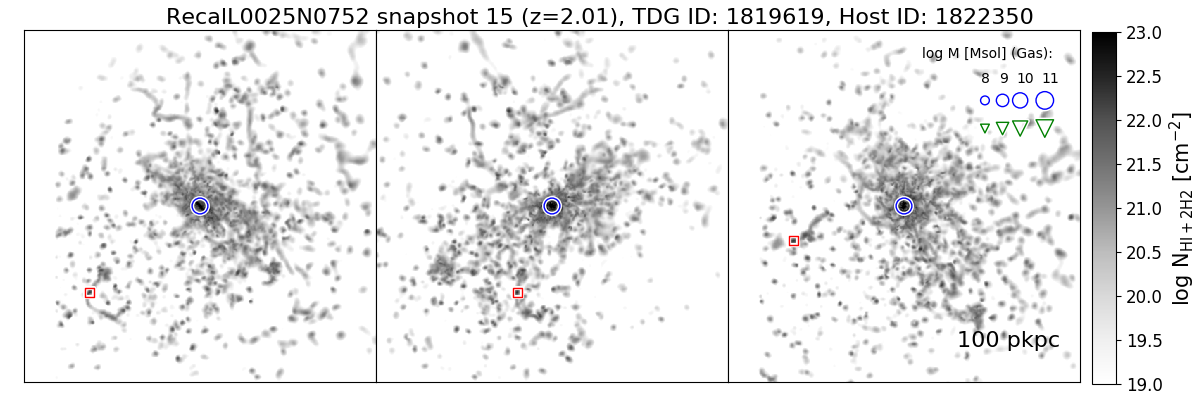}
		\includegraphics*[width=0.75\linewidth]{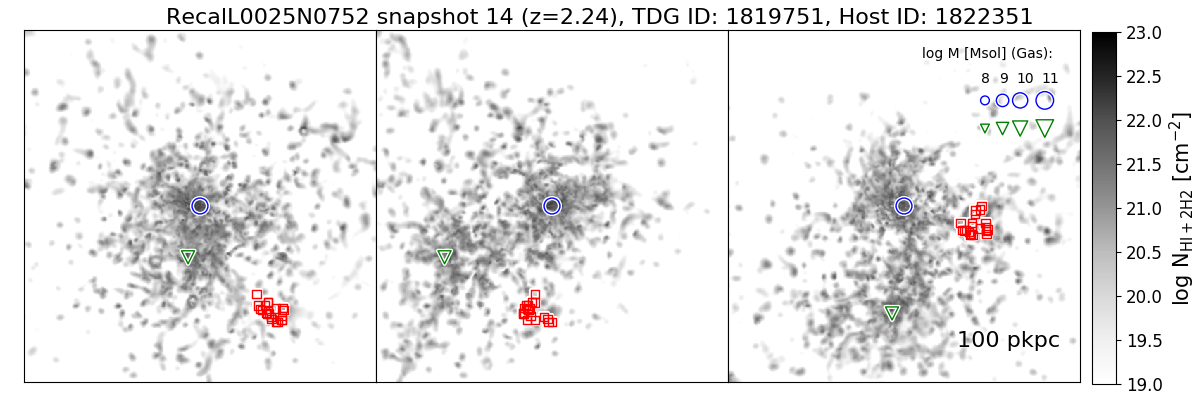}
		\includegraphics*[width=0.75\linewidth]{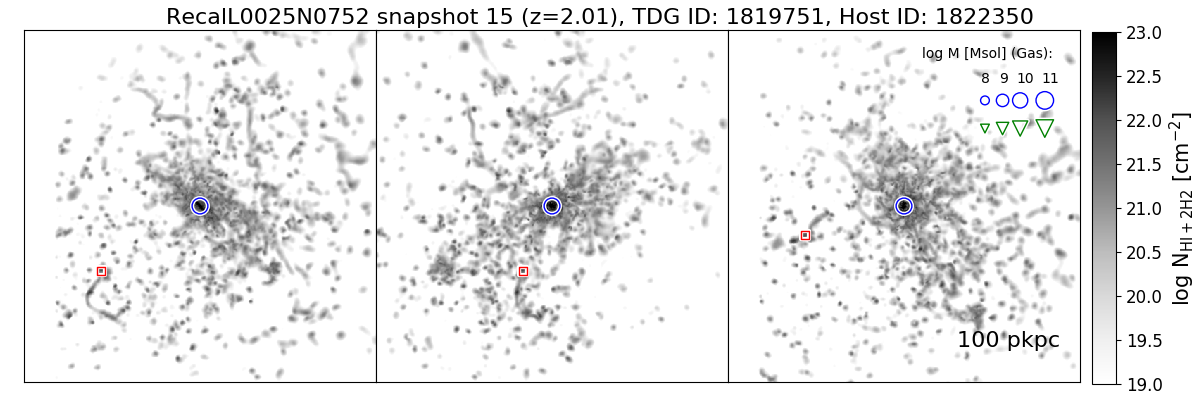}
		\caption{As Fig.~\ref{fig:TDGC1} but for Recal-2 (top two panels) and Recal-3 (bottom two panels). The closely separated TDGCs originate from the same host galaxy that is in a close encounter at snapshot 14. Their metallicities are clearly above the mean mass-metallicity relation for dwarf galaxies at this redshift (see Fig.~\ref{fig:MZrelation}). Note that the difference in lookback time between snapshots 14 ($z=2.24$) and 15 ($z=2.01$) is only around 330 Myr. }
		\label{fig:TDGCRecal45}
	\end{center}
\end{figure*}

\begin{figure*} 
	\begin{center}
		\includegraphics*[width=0.75\linewidth]{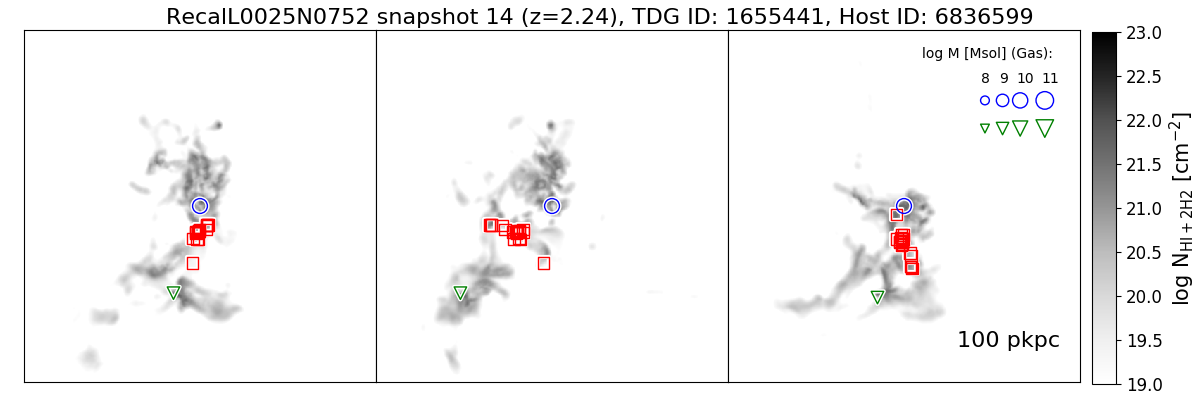}
		\includegraphics*[width=0.75\linewidth]{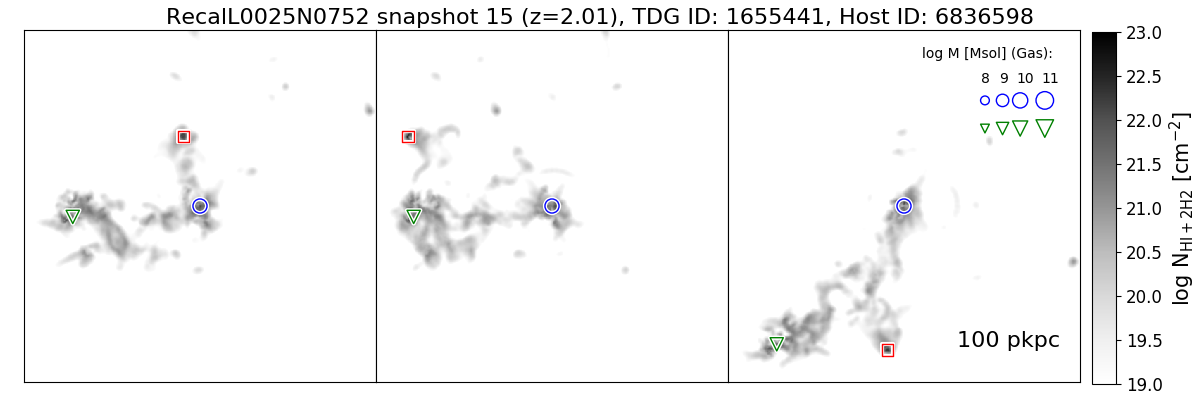}
		\caption{As Fig.~\ref{fig:TDGC1} but for TDGC Recal-4. While the stellar metallicity is not clearly elevated (Fig.~\ref{fig:MZrelation}), the TDG material originates from a close interaction and forms a gravitationally bound structure between redshifts 2.24 (snapshot 14) and 2 (snapshot 15).}
		\label{fig:TDGCRecal6}
	\end{center}
\end{figure*}

\section{Rejected candidates}\label{sec:rejected}

\begin{figure*} 
	\begin{center}
		\includegraphics*[width=0.75\linewidth]{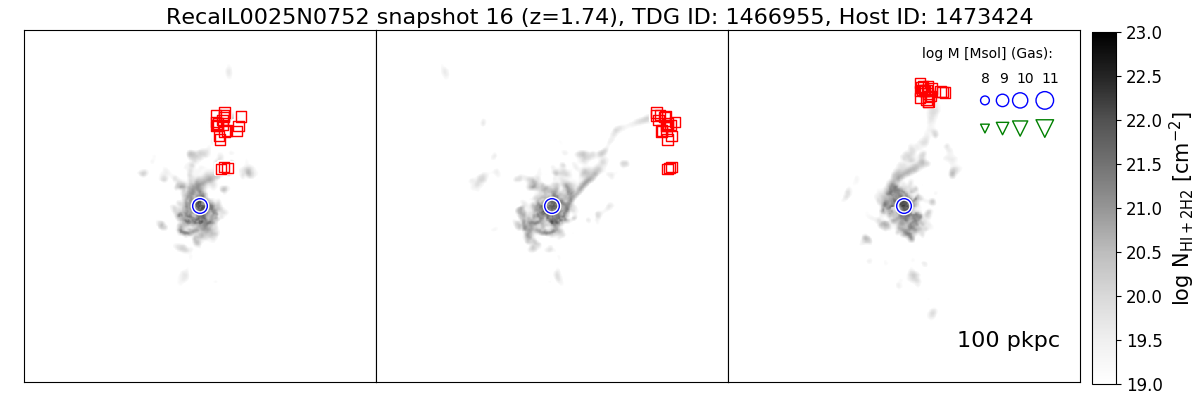}
		\includegraphics*[width=0.75\linewidth]{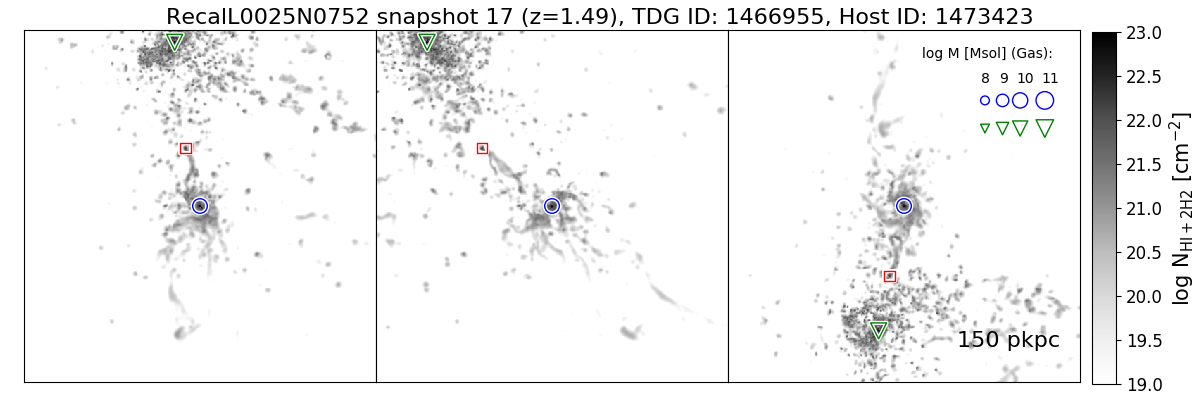}	
		\caption{As Fig.~\ref{fig:TDGC1} but for Rej-1. Visually, this candidate looks like it has a tidal origin: the traced back particles are located in the outskirts of the host galaxy and the TDGC forms within the bridge between two interacting galaxies. However, the extremely low stellar metallicity of the candidate ($\log_{10} Z/\Zsol < -3$) compared to the much higher stellar metallicity of the host galaxy at both snapshots ($\log_{10} Z/\Zsol = -0.21$ at $z = 1.49$ and $\log_{10} Z/\Zsol = -0.33$ at $z = 1.74$) hint at a misidentification of an in-falling structure.}
		\label{fig:TDGCRecal1}
	\end{center}
\end{figure*}

\begin{figure*} 
	\begin{center}
		\includegraphics*[width=0.75\linewidth]{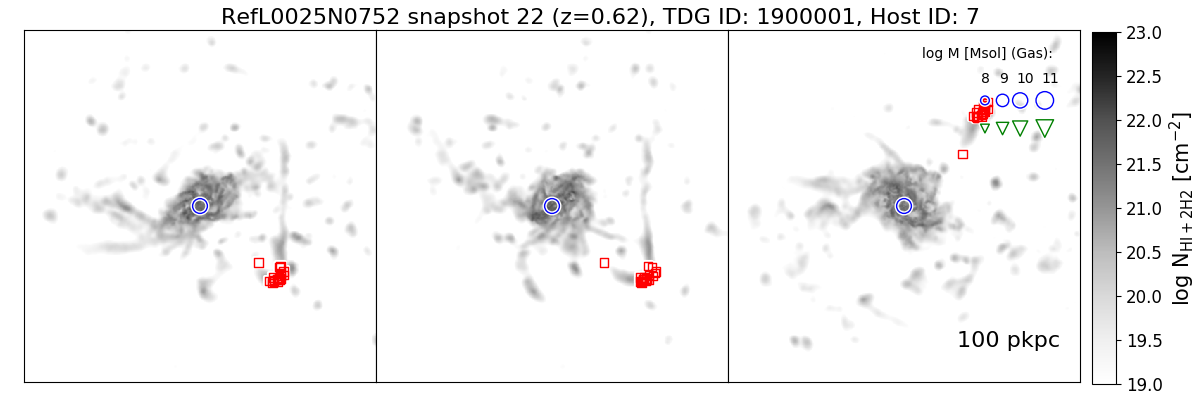}
		\includegraphics*[width=0.75\linewidth]{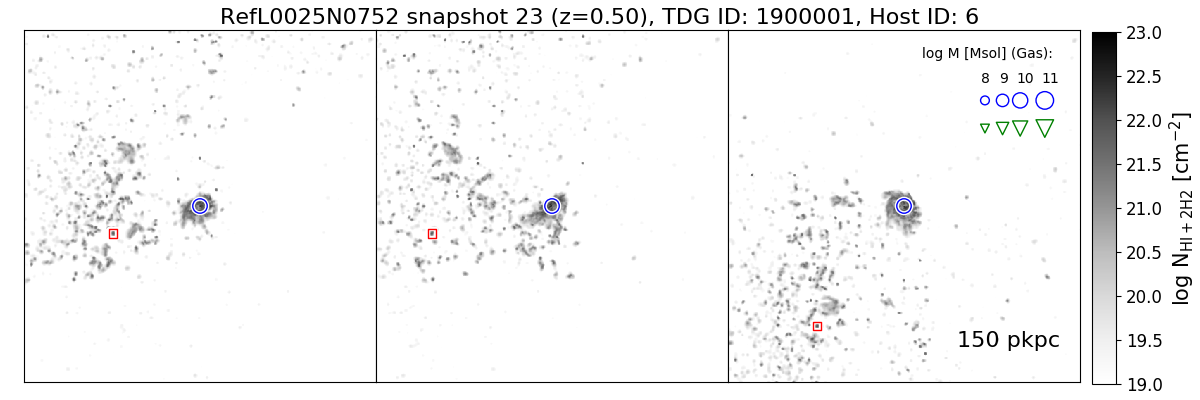}
		\caption{As Fig.~\ref{fig:TDGC1} but for Rej-2. The traced-back particles are located in the outskirts of the host galaxy. No neighbouring galaxy is located within 150 ckpc of the host galaxy at both considered snapshots. Based on the morphology and the lack of galaxy interaction this is a candidate for a self-gravitating object that formed out of ram pressure stripped gas, even though its stellar metallicity is lower than the mean stellar metallicity of dwarf galaxies of similar stellar masses. It is therefore excluded from our TDG sample but still an interesting example of a ``fireball" \citep{yoshida_strange_2008}.}	
		\label{fig:TDGC5}
	\end{center}
\end{figure*}


\bsp	
\label{lastpage}
\end{document}